\documentclass[amsmath,amssymb,superscriptaddress,showkeys,10pt,nofootinbib, twocolumn]{revtex4-2}

\usepackage[utf8]{inputenc}
\usepackage{booktabs}
\usepackage{array, multirow}
\usepackage{csquotes}
\usepackage{subfloat}
\usepackage{textcomp}
\usepackage{graphicx}
\usepackage{dcolumn}
\usepackage{bm}
\usepackage[title]{appendix}
\usepackage{hyperref}
\usepackage{color}
\usepackage[normalem]{ulem}
\usepackage{enumerate}
\usepackage[shortlabels]{enumitem}
\usepackage{amsmath,amssymb,amsfonts,amsthm}
\newtheorem*{definition}{Definition}

\newcommand{\tcr}[1]{{#1}}
\newcommand{\tcb}[1]{{#1}}
\newcommand{\tcg}[1]{{#1}}

\newcommand{\Eth}{E_{\rm th}}
\newcommand{\Ipeak}{I_{\rm peak}}

\newcommand{\Taupeak}{\tau_{\rm peak}}

\DeclareMathOperator{\sign}{sign}

\def\copyrightname{Copyright (2025) E.~G. Evstatiev, M.~H. Hess, N.~D. Hamlin, and B.~T. Hutsel.
  This article is distributed under a Creative Commons Attribution-NonCommercial-NoDerivs
  4.0 International (CC BY-NC-ND) License. This article appeared in Phys. Plasmas 32, 062707 (2025)
  and may be found at \url{https://doi.org/10.1063/5.0254084}.}

\begin{document}

\title{Non-local, diamagnetic electromagnetic effects in magnetically
insulated transmission lines}
\author{E.~G. Evstatiev}
\email{egevsta@sandia.gov (correposnding author)}
\author{M.~H. Hess}
\author{N.~D. Hamlin}
\author{B.~T. Hutsel}
\affiliation{Sandia National Laboratories, Albuquerque, New Mexico 87185}

\date{17 June, 2025}

\begin{abstract}
  We identify\footnote{\copyrightname} the time-dependent physics responsible for the critical reduction of current
  losses in magnetically insulated transmission lines (MITLs) due to
  uninsulated space charge limited (SCL) currents of electrons emitted by 
  field stress. A drive current of sufficiently short pulse length introduces a
  strong enough time dependence that steady state results alone become inadequate for
  the complete understanding of current losses.
  The time-dependent physics can be described as a non-local, diamagnetic electromagnetic response of
  space charge limited currents. As the pulse length is increased or equivalently, the MITL length reduced,
  these time-dependent effects diminish and current losses converge to those predicted
  by the well-known Child-Langmuir law in the external (vacuum) fields.
  We present a simple one-dimensional (1D) model that encapsulates the essence of this physics.
  We find excellent agreement with 2D particle-in-cell (PIC) simulations for two MITL geometries,
  Cartesian parallel plate and azimuthally symmetric straight coaxial.
  Based on the 1D model, we explore various scaling dependencies of MITL losses
  with relevant parameters, e.g., peak current, pulse length, geometrical dimensions, etc.
  We propose an improved physics model of magnetic insulation in the form of a Hull curve,
  which could also help improve predictions of current losses by common circuit element codes, such as BERTHA.
  Lastly, we describe how to calculate temperature rise due to electron impact within the 1D model.
  
\end{abstract}


\maketitle

\section{Introduction}
\label{sec:introduction}

Pulsed power devices \cite{sinars_review_2020,McBride:2018,stygar_design_2015} require the transport
of large amounts of power over relatively long distances. Previous experiments
\cite{baranchikov_transfer_1977,di_capua_propagation_1979,vandevender_long_1979}
have shown that $80$\% or more of the input energy
can be delivered to a load up to $10\,$m away from a generator. At Sandia National Laboratories'
$Z$ machine, up to $40\,$terawatts of power is delivered from the stack to a load over more than a meter
distance, with little losses. At the heart of the ability to transport such large amounts
of power is the physics of magnetic insulation (also called self-insulation)
in magnetically insulated transmission lines (MITLs)
\cite{hull_magnetic_insulation_1921,lovelace_theory_1974,Creedon:1975,di_capua_magnetic_1983,Mendel:1983,Ottinger:2006}.
 
\tcg{During the beginning of the current pulse, electron emission from the cathode is observed when 
the electric fields exceed values of about $20$--$30\,$MV/m \cite{di_capua_propagation_1979}.}
These electrons may or may not be initially insulated, something highly dependent on MITL geometry
and temporal current pulse profile.
For the prototypical sine squared current pulse shape used throughout this work,
\footnote{The conclusions in this work also hold for more realistic current pulses used on $Z$,
however, we do not show results for such pulses.}
uninsulated electrons are typically observed in \textit{long} MITLs (see the \tcb{definition} below).
Uninsulated electrons are a major contributor to current loss and anode temperature rise,
and are for this reason a major concern when designing long MITLs.
\tcr{Additionally, when the anode temperature rises above approximately 400\,$^\circ\!\!$~C,
  contaminants are thermally desorbed from the electrode surface \cite{sanford_thermal_desorption_1989},
  becoming a source of ions at the anode. Ion Larmor radii are typically larger than those of electrons and for this reason,
  in many situations ions can account for a significant fraction of the current loss.
  Ion current losses are not treated in this work but will be considered in the future.}

Current losses are being predicted in one of two ways.
First, fully kinetic electromagnetic (EM) particle-in-cell (PIC) simulations are
considered as the most general and trusted approach
\cite{pointon_particle_cell_2007,pointon_current_2009,ottinger_rescaling_2006,
  welch_electrode_2019,bennett_current_2019,welch_fast_2020}.
By its nature, the PIC method \cite{birdsall_plasma_2005,hockney_computer_2021} is
very computationally expensive. For this reason, approximate one-dimensional (1D)
circuit element models (CEM) have been devised
\cite{hinshelwood_bertha_1983,huttlin_tlines_1983,kiefer_screamer_1985,weseloh_tlcode_1989,harjes_circuit_2003,spielman_design_2019}.
Since CEM models are only 1D, their computational cost is many orders of magnitude
lower than that of PIC, and for this reason are preferred for preliminary MITL design, leaving
PIC simulations for final verification. 
Both methods have shown reliable predictions of current losses in time-dependent settings
(i.e., using current pulses), implying that they both capture the essential physics.

In this work we propose an alternative one dimensional method based on electromagnetic fields
and sources instead of circuit elements. One may consider such a formulation as the basis for
the circuit element method. Working with the fields and sources
has the advantages that (i) one can interpret the results in a physically intuitive way;
and (ii) a more direct comparison with electromagnetic PIC simulations is possible.
The latter is important in the verification of preliminary MITL designs.

The main assumptions of our one dimensional model are these:
\begin{enumerate}[label={(A-\roman*)}] 
  \itemsep0em
  
\item Infinitely small anode-cathode (AK) gap; i.e., AK gap that is much smaller
  than all other MITL dimensions. This assumption allows us
  to decouple the 2D problem into two 1D problems.
  It also assumes that changes of (local in space) space charge limited (SCL)
  currents are instantaneous, i.e., we neglect the time for electrons to cross the AK gap;
   \label{assumption:AK_gap}

\item SCL currents vary according to the well known \tcg{non-relativistic} Child-Langmuir law
  \cite{child_discharge_1911,langmuir_effect_1913};
  \label{assumption:CL_current}

\item A magnetic insulation model, in which SCL currents ramp down smoothly as the magnetic field
  increases, according to what we refer to as a \textit{Hull curve} (see Sec.~\ref{sec:hull_curve}).
\label{assumption:Hull_curve}


\end{enumerate}

The ``standard'' definition of a \textit{long} MITL is one in which an electromagnetic
wave traverses its length in time comparable to the length of the electromagnetic (current)
pulse itself.   
However, based on the results in \tcb{this work}, we propose an alternative definition of a long MITL,
which more accurately accounts for the electromagnetic effects and
more generally, the effects of time dependence. 
\begin{definition}
  A long MITL is one in which time-dependent 
  processes result in a significant deviation of quantities of merit
  from their steady state values.
\end{definition}
\noindent
Relevant quantities of merit include fields, current and charge densities,
loss current and charge, anode temperature rise, etc. An example considered
in this work is a $100\,$ns pulse launched into a $1\,$m-long MITL.
Although the travel time (back and forth) of the EM wave in this MITL is only $6.67\,$ns,
much smaller than the pulse length of $100\,$ns, we find that current losses are greatly
overestimated by using steady-state theory alone. This and many similar
results have served as the motivation for the above definition.

In the main exposition, we consider a parallel plate Cartesian geometry,
which allows for a certain degree of analytical treatment.
The straight coaxial geometry with a finite AK gap can also be treated analytically;
however, its treatment greatly simplifies in the small AK gap limit, for which
we outline some detail in Appendix~\ref{sec:coaxial_MITL}.
In general geometries, we expect the physical nature of the processes
to be the same, albeit with quantitative differences.

The paper is structured as follows. Section~\ref{sec:introduction} is an introduction.
Section~\ref{sec:1D_model} introduces the reduced physics 1D electromagnetic model.
Section~\ref{sec:diamagnetic} describes the non-local, diamagnetic time-dependent
electromagnetic effects in long MITLs. Section~\ref{sec:scalings} presents scaling of MITL
losses with various parameters. Section~\ref{sec:hull_curve} presents a
model for magnetic insulation, improving upon those found in present CEM models.
Section~\ref{sec:temperature} presents the temperature rise diagnostics
within the 1D model. Section~\ref{sec:conclusions} summarizes and presents final concluding remarks.

\section{A one-dimensional electromagnetic model of a {MITL}}
\label{sec:1D_model}

In the geometry under consideration, all electromagnetic waves,
external and emitted by SCL currents, propagate as TEM modes.
We choose  $z$ as the propagation direction, $x$ as the transverse (AK gap) direction,
then $y$ is the ignorable direction (symmetry dimension).
Correspondingly, the non-trivial components of the fields and currents (current densities)
are $E_x$, $B_y$, $j_x$. Quantities also depend on time, $t$.
Waves are launched at $z=0$ in the positive $z$-direction. Waves propagating
in the negative $z$-direction (to the left) are absorbed at $z=0$.
There is a perfectly conducting load (short) at the opposite end of the MITL, $z=L$,
so that waves are reflected off that end.

The basic model of SCL emission assumes that a local SCL
current is ``once-on-always-on.'' In other words, we assume that once the threshold
for SCL emission is exceeded at a particular location and instant of time,
the inventory of plasma is sufficient to supply any amount of electrons necessary
to sustain zero electric field at the cathode at maximal SCL current;
this is the typical assumption in PIC simulations as well.
Some models assume initially a source-limited emission, which
later transitions into SCL emission. We do not include such detail 
as it does not qualitatively affect the following discussion and results.

As already mentioned, the small AK gap assumption \ref{assumption:AK_gap} allows us to decouple the
two spatial dimensions of the problem. The two 1D problems consist of an electromagnetic
problem along the direction of EM wave propagation and an electrostatic problem in the direction across
the AK gap. First, let us consider the electromagnetic part of the model.
Suppose a spatial location $z_0$ starts emitting an SCL current at time $t_0=0$.
This emission model can be mathematically represented as a Dirac
delta-function in space and a Heaviside step-function in time
%
\begin{equation}
  j_x(z,t) = I_0\, \delta(z-z_0)\, \theta(t),
  \label{eq:point_source}
\end{equation}
with current $I_0$ per unit length (unit length in the ignorable $y$-direction).
We assume $I_0$ to be the current \textit{on the conductor}. The relation between
the current in the AK gap and the current on the conductor is illustrated
in Fig.~\ref{fig:diamagnetic}. 
We also note that the delta function has dimensions of inverse length
while the step function is dimensionless; thus $j_x$ has the dimensions of current
density, e.g., \tcb{A/m$^2$} in SI units.

Working in the Coulomb gauge (vector potential satisfying $\nabla\cdot\mathbf{A}=0$),
the equation for the electromagnetic waves can be reduced to 
\begin{equation}
  \frac{\partial^2 A_x(z,t)}{\partial z^2} - \frac{1}{c^2}\frac{\partial^2 A_x(z,t)}{\partial t^2} = -\mu_0 j_x(z,t)
  \label{eq:wave_eq}
\end{equation}
with electric and magnetic fields given by 
\begin{align}
  E_x(z,t) &= -\frac{\partial A_x(z,t)}{\partial t}, \label{eq:E_x_def}\\
  B_y(z,t) &= \left(\nabla\times \mathbf{A}\right)_y = \frac{\partial A_x(z,t)}{\partial z}.\label{eq:B_y_def}
\end{align}

Using the fundamental solution (Green's function) of the wave operator (d'Alembertian),
$\partial_t^2 - c^2\partial_x^2$, in one spatial dimension (of infinite $z$-extent),
%
${\cal E}_1 = \frac{1}{2c} \theta\left(t - \frac{|z|}{c}\right)$,
%
we can follow the general rule for finding the special solution of Eq.~\eqref{eq:wave_eq}
by convolution \cite{vladimirov_equations_1984}, $A_x(z,t) =  {\cal E}_1\star \left(c^2\mu_0 j_x\right)$.
For the current source \eqref{eq:point_source}, the integrals are easy to evaluate and give
\begin{align}
  A_x(z,t) =  \frac{I_0Z_0}{2} \left(t - \frac{|z-z_0|}{c}\right)\, \theta\left( t - \frac{|z-z_0|}{c}\right),
             \label{eq:convolution}
\end{align}
where $\epsilon_0$ and $\mu_0$ are the permittivity and permeability of vacuum,
$Z_0=\sqrt{\mu_0/\epsilon_0}\approx 376.7\,\Omega$ is the vacuum (free space) impedance,
and $c$ is the speed of light in vacuum.
The fields following from Eqs.~\eqref{eq:E_x_def}--\eqref{eq:convolution} are
\begin{align}
  E_x(z,t) &= - \frac{I_0 Z_0}{2} \, \theta\left( t - \frac{|z-z_0|}{c}\right),\label{eq:E_x}\\
  B_y(z,t) &=  \sign(z-z_0)\, \frac{E_x(z,t)}{c}\label{eq:B_y}.
\end{align}
Note that the magnetic field has a discontinuity at the location, $z=z_0$,
since this is a surface (sheet) current. The jump across the discontinuity is the
difference between the values on the right and on the left of the current sheet,
and gives the expected result
\begin{equation}
  \left[B_y\right]_{z=z_0} = \left| B_y(0^+)-B_y(0^-) \right| = \frac{I_0Z_0}{c} = \mu_0 I_0.
  \label{eq:B_y_jump}
\end{equation}
%

An external (vacuum) field can be launched into the MITL by setting up a current source at the launch side of the
MITL, $z_0=0$, with a desired temporal dependence. For example, a sine squared current pulse
peaking at value $\Ipeak$ at time $t=\Taupeak$ can be set up by
\begin{equation}
  j_x^{\rm ext}(z,t) = \Ipeak \, \delta(z)\, \sin^2\left(\frac{\pi t}{2\Taupeak}\right).
  \label{eq:launch_ext_waves}
\end{equation}
The corresponding fields can be calculated by following the same
steps leading to Eq.~\eqref{eq:convolution}:
\begin{align}
  E^{\rm ext}_x(z,t)
  &= - \frac{\Ipeak Z_0}{2} \, \sin^2\left(\frac{\pi( t-|z|/c)}{2\Taupeak}\right),
    \label{eq:E_x_sin2}\\
  B^{\rm ext}_y(z,t) &= \frac{\tcb{E^{\rm ext}_x(z,t)}}{c}\label{eq:B_y_sin2},
\end{align}
where the magnetic field sign corresponds to a wave launched in the
positive $z$-direction (see Fig.~\ref{fig:diamagnetic}).
Our numerical implementation indeed uses Eq.~\eqref{eq:launch_ext_waves} to launch
external waves into the MITL.

Now consider the dynamics within the AK gap. Assuming the (vacuum) electric field at the cathode
has exceeded the SCL emission threshold, at any instant of time
we assume that the SCL current is given by Child-Langmuir's law,
\begin{equation}
  \tcb{j_{\rm SCL}(z,t) = \frac{4\epsilon_0}{9 d^2}\sqrt{\frac{2 e}{m}} V(z,t)^{3/2}},
  \label{eq:j_CL}
\end{equation}
where $d$ is the size of the AK gap, $e$ and $m$ are the charge and mass of the electron.
\tcb{$V(z,t)$ is related to the electric field in \tcb{Eq.~\eqref{eq:E_x_def}} as follows.
  We identify the value of the electric potential at the anode at any $(z,t)$ with
  the voltage at the anode, \tcb{$V_0 \equiv V(z,t) = -\int_0^d\!dx\,E_x(z,t) = - E_x(z,t)\, d$,
    where $E_x(z,t)$ is given by Eq.~\eqref{eq:E_x_def},
    $x=0$ is the cathode and $x=d$ is the anode.}
  This gives the potential determining the value of
  the Child-Langmuir current
  %
\begin{equation}
  \tcb{j_{\rm SCL}(z,t) = \frac{4\epsilon_0}{9 d^2}\sqrt{\frac{2 e}{m}} \left[E_x(z,t)\,d\right]^{3/2}.}
  \label{eq:j_CL2}
\end{equation}
(Absolute values are taken whenever necessary.)}
\tcb{Eq.~\eqref{eq:j_CL2} is seen as Assumption~\ref{assumption:CL_current}.
  The final expression for the current density is obtained by applying Assumption~\ref{assumption:Hull_curve}
  with the help of an analytic fit of PIC simulation data, as discussed in Sec.~\ref{sec:hull_curve}.
  Thus, the current density entering the right-hand side of Eq.~\eqref{eq:wave_eq} is given by}
\begin{equation}
  \tcb{j_x(z,t) = Y\!\left(\frac{B_y(z,t)}{B_{\rm crit}(z,t)}\right)\, j_{\rm SCL}(z,t),}
  \label{eq:j_x_hull_correction}
\end{equation}
\tcb{where the function $Y$ is given in Eq.~\eqref{eq:Hull_fit},
  $B_y$ is given by \eqref{eq:B_y_def}, and $B_{\rm crit }$ is given by \eqref{eq:Lovelace_Ott}.
  The system of equations \eqref{eq:wave_eq}--\eqref{eq:B_y_def}, \eqref{eq:j_CL2}, \eqref{eq:j_x_hull_correction}
  is complete. We have shown explicitly all arguments in Eq.~\eqref{eq:j_x_hull_correction}
  in order to emphasize the fact that all quantities defining our 1D model
  are local in space and time. It is also assumed that when the electric field at the cathode surface is positive,
  i.e., has the ``wrong'' sign, no SCL current is emitted.
}

In the presence of \textit{uninsulated} SCL emission at $(z,t)$, the electric field within the AK gap
is modified and  is no longer constant. The Child-Langmuir condition of zero electric
field at the cathode is satisfied by (e.g., see \cite{davidson_physics_2001})
\begin{equation}
  V(x; z,t) = V_0 \left(\frac{x}{d}\right)^{4/3}.
  \label{eq:V_CL}
\end{equation}
The semicolon in the list of
arguments in Eq.~\eqref{eq:V_CL} emphasizes the fact that $z$ and $t$ are here
considered as parameters, not actual arguments to $V$.
\tcb{The modified electric field\,---\,denote that by $\widetilde{E}_x(x;z,t)$\,---\,is
  found from \eqref{eq:V_CL} and at the anode has the value
  (for definiteness, we compare fields and currents between models near the anode)}
\begin{equation*}
\widetilde{E}_x(x=d) = -\left. \frac{dV(x)}{dx}\right|_{x=d} = -\frac{4}{3}\frac{V_0}{d}.
\end{equation*}
Therefore, the electric field at the anode \tcb{in the presence of uninsulated SCL emission} is given by
\begin{equation}
  \widetilde{E}_{x}(x=d;z,t) = \frac{4}{3}E_x(z,t),
  \label{eq:E_tot}
\end{equation}
from which
\begin{equation}
  \tcb{V_0 = V(z,t) = \frac{3}{4} \widetilde{E}_x(z,t)\, d.}
  \label{eq:V_0}
\end{equation}
We use equations \eqref{eq:E_tot} and \eqref{eq:V_0} in two cases.
\tcb{First, Eq.~\eqref{eq:E_tot} allows us to find the electric field at the
  anode in the case of uninsulated SCL emission 
  from the solution of the 1D model, $E_x(z,t)$,
  by simply multiplying it by $4/3$; conversely, this factor is not
  included at any $(z,t)$ prior to initiation of SCL emission 
  or after magnetic insulation has established. The electric field $\widetilde{E}_x(z,t)$
  can be directly compared to its counterpart from a full electromagnetic PIC simulation.
  Keep in mind that this is a dynamic system and at the same location $z$,
  even after SCL emission has once started, magnetic insulation can repeatedly
  cease and at a later time reestablish.}  
And second, relation \eqref{eq:V_0} is instrumental to correctly
determine the Hull curve from PIC simulations,
as discussed in Sec.~\ref{sec:hull_curve}.
\tcb{For the remainder of the paper, we will denote the electric field in the system
  by $E_x(z,t)$, implicitly including the factor $4/3$ whenever necessary.}

\subsection*{Numerical considerations}

We note that
the simulation of SCL emission by PIC has both physical and numerical aspects. The physical
aspect is that of the Child-Langmuir law. In emission due to field stress, i.e.,
after a certain threshold value of the (normal) electric field has been exceeded,
electrons are emitted from the cathode and conduct space charge limited
current in the anode-cathode gap, when uninsulated.
The numerical aspect aims to implement an algorithm that uses computational particles
and finite time steps, that is numerically stable, and that produces results
in agreement with the physical aspect. In this regard, ``numerical knobs''
exist in PIC codes which, if not used carefully, may noticeably distort the results.
For example, Sandia's PIC code EMPIRE \cite{bettencourt_empire-pic_2021},
implements a function that ramps up SCL current in time from a small fraction
to its full value, according to some temporal dependence. A similar
capability exists in CHICAGO
\cite{welch_implementation_2004,welch_adaptive_2007,genoni_fast_2010,welch_fast_2020}.
We believe that most PIC codes allow for similar numerical knobs.
Accordingly, we have implemented a similar parameter \tcb{in} our 1D electromagnetic model,
allowing for a closer comparison of the results. Typical ramp times used in our
simulations were in the $0.25$--$0.5\,$ns range.
We refer the reader to the extensive literature for more detail
(e.g., \cite{birdsall_plasma_2005,pointon_slanted_1991,watrous_improved_2001,luginsland_beyond_2002},
with a brief discussion also given in Ref.~\cite{evstatiev_efficient_2023}).

Simulations with the 1D model were done with resolutions in the range $10$--$100\,\mu$m,
with the smaller resolution used in the scaling studies in Sec.~\ref{sec:scalings}
as well as in checks for numerical convergence. The time step in the 1D simulations was
determined by a Courant-Friedrichs-Lewy (CFL) condition of $0.6$.
The PIC simulations with CHICAGO and EMPIRE were
done with resolutions $100$--$200\,\mu$m. The time step was determined
by the more restrictive condition of resolving the cyclotron period with
about $20$--$30$ points; a typical value was $dt=5\times10^{-14}\,$s.


\section{Non-local, diamagnetic electromagnetic effects of {SCL} currents}
\label{sec:diamagnetic}

\begin{figure}[t!!]
  \centering
\includegraphics[width=0.33\textwidth]{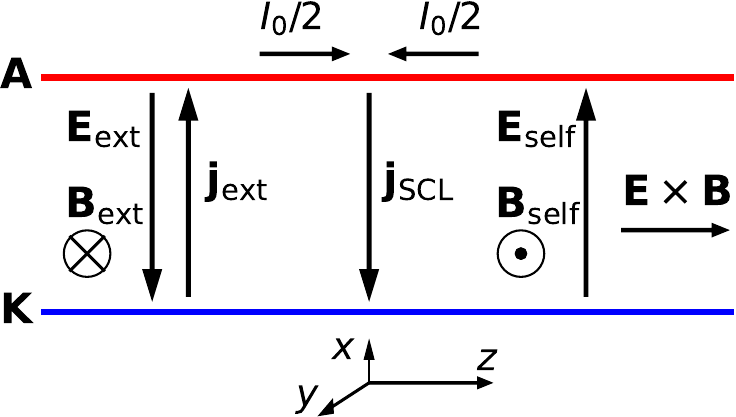}
\caption{Directions of fields and currents within the MITL
  and on the conductor. Currents with subscript '0' denote current on the conductor.
  The field directions are shown only for waves propagating in the positive $z$-direction;
  waves propagating to the left have a reversed magnetic field direction.
  \tcb{The fields labeled ``self'' are emitted from the
    point source current  $\mathbf{j}_{\rm SCL}$ [cf. Eq.~\eqref{eq:point_source}] whereas the
    fields labeled ``ext'' are launched by an external (generator) source
    [cf. Eqs.~\eqref{eq:launch_ext_waves}--\eqref{eq:B_y_sin2}].
  The self-waves in both directions are thus diamagnetic.}}
\label{fig:diamagnetic}
\end{figure}

We observe from Eqs.~\eqref{eq:E_x}, \eqref{eq:B_y} that in a long MITL,
a wave emitted by a local SCL current per the
``once-on-always-on'' rule propagates throughout the full length of the MITL with
\textit{undiminished amplitude}, i.e., is \textit{non-local}. The \textit{diamagnetic} nature of the waves
can be seen from Fig.~\ref{fig:diamagnetic}.
\tcb{The fields labeled ``self'' are emitted by SCL currents, while those labeled ``ext''
  are the fields launched externally into the MITL (by a generator).
  The figure also shows
  the corresponding currents on the electrodes (conductors).
  In the rest of the paper we omit the ``self'' and ``ext'' subscripts when we see no confusion;
  we also use the terms ``external'' and ``vacuum'' fields interchangeably.}
We discuss these two effects next.

\begin{figure}[t!!]
  \centering
  \includegraphics[width=0.4\textwidth]{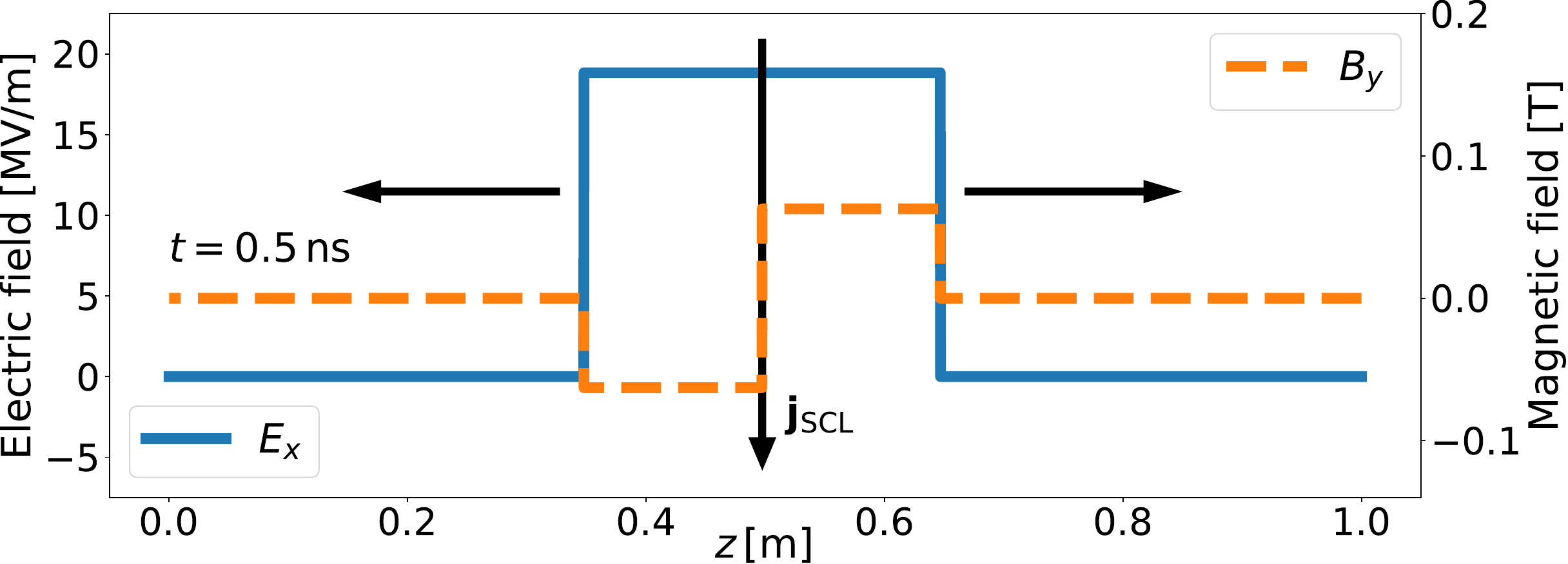}\vspace{5pt}\\
  \includegraphics[width=0.4\textwidth]{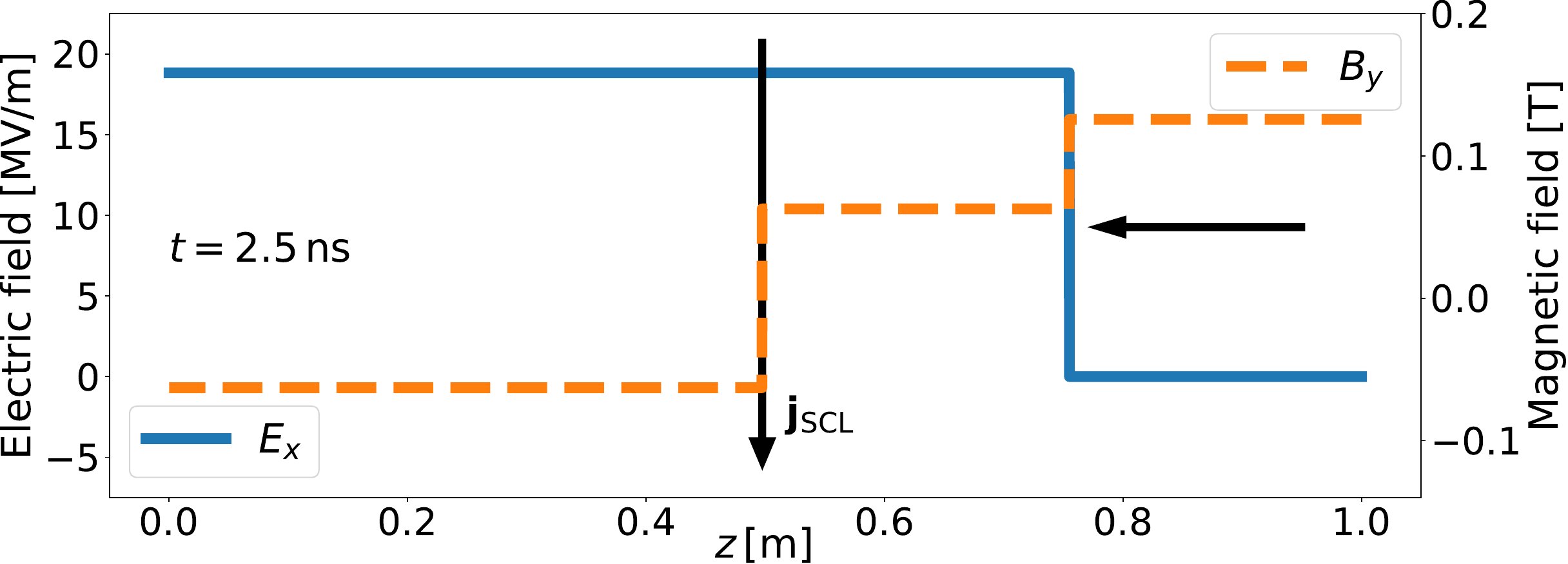}\vspace{5pt}\\
  \includegraphics[width=0.4\textwidth]{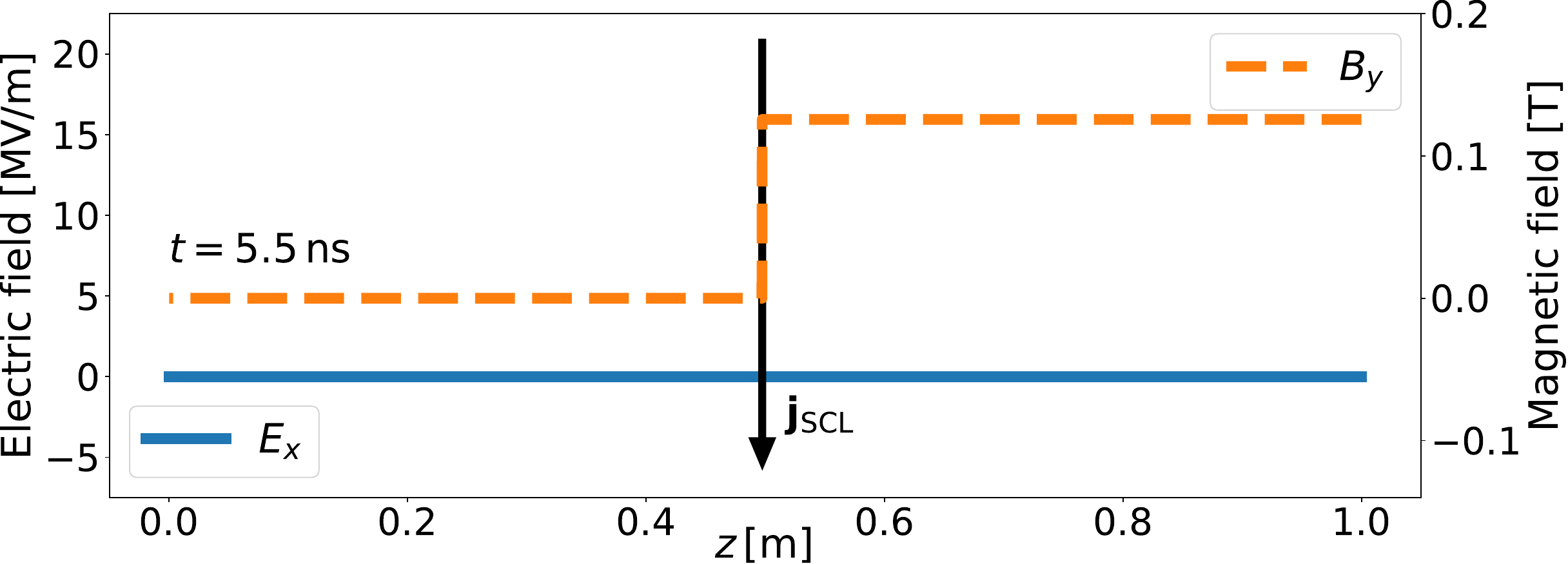}\\  
  \caption{Time evolution of a wave emitted from a point source with a step function time dependence, Eq.~\eqref{eq:point_source}.
    Top panel: two waves propagating away from the current source and before reflection from the load.
    Middle panel: the right-propagating wave after reflection is now propagating to the left, the left-propagating
    wave has exited the system. Bottom panel: steady state after the EM waves have left the MITL.}
\label{fig:single_wave}
\end{figure}

The non-local effect of SCL currents is demonstrated in Fig.~\ref{fig:single_wave}, where the temporal evolution
of a wave from a point current source at $z_0=50\,$cm with $I_0=100\,$kA/m is shown [cf. Eq.~\eqref{eq:point_source}].
The current source indicated in the figure is in the negative $x$-direction, hence, the
fields have opposite signs to those in Eqs.~\eqref{eq:E_x}, \eqref{eq:B_y}.
The MITL has a length of $L=1\,$m and an AK gap $d=5\,$mm.
At $t=0.5\,$ns, two waves are seen propagating away from the source (directions are indicated
by arrows) and before reflection from the short at $z=1\,$m. At $t=2.5\,$ns, the right-propagating
wave has reflected and is now propagating to the left, while the left-propagating wave has exited the MITL.
It is seen that since the amplitudes of the waves due to this source are non-diminishing in time
and in space (up to the wave front), the electric field of the emitted
(right-propagating) wave cancels out exactly the electric field of the reflected
(now left-propagating) wave so that the total electric field behind the reflected wave is exactly zero.
The magnetic fields of the incoming and outgoing waves \textit{add}, and 
at $t=5.5\,$ns, a steady state has been established with only a constant magnetic field present
in the MITL. \tcb{The magnitude of this magnetic field is zero for $z<0.5\,$m and
  jumps to $\left[B_y\right]_{z=50\,\mbox{cm}} \simeq 0.126\,$T for $z>0.5\,$m, according to Eq.~\eqref{eq:B_y_jump}.}
In a time-dependent system (time-dependent current source),
the emitted and reflected waves would not cancel exactly due to the temporal
delay between the time of emission and the time the reflected wave has traveled
back to the source, during which the emitted wave from the source will have changed its amplitude.
Note that this relatively low current
(peak currents of interest in the $Z$ machine are $20$--$30\,$MA, while smaller pulsed power
machines have currents of order mega-ampere) already generates an electric field with amplitude $\simeq 19\,$MV/m,
comparable to the threshold for SCL emission.
Conversely, currents of this magnitude may be expected during the beginning of SCL emission.

\begin{figure}[t!!]
  \centering
  \includegraphics[width=0.4\textwidth]{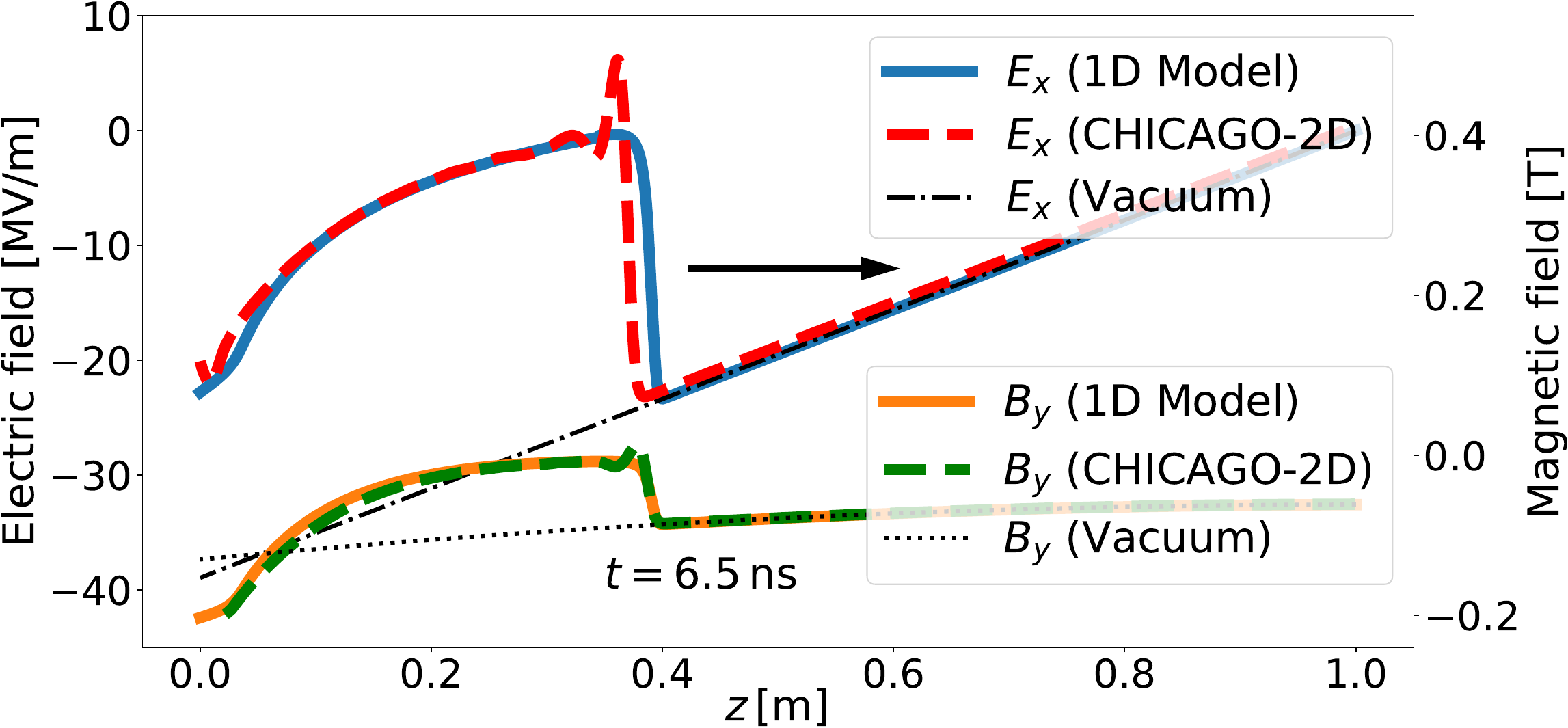}\vspace{5pt}\\
  \includegraphics[width=0.4\textwidth]{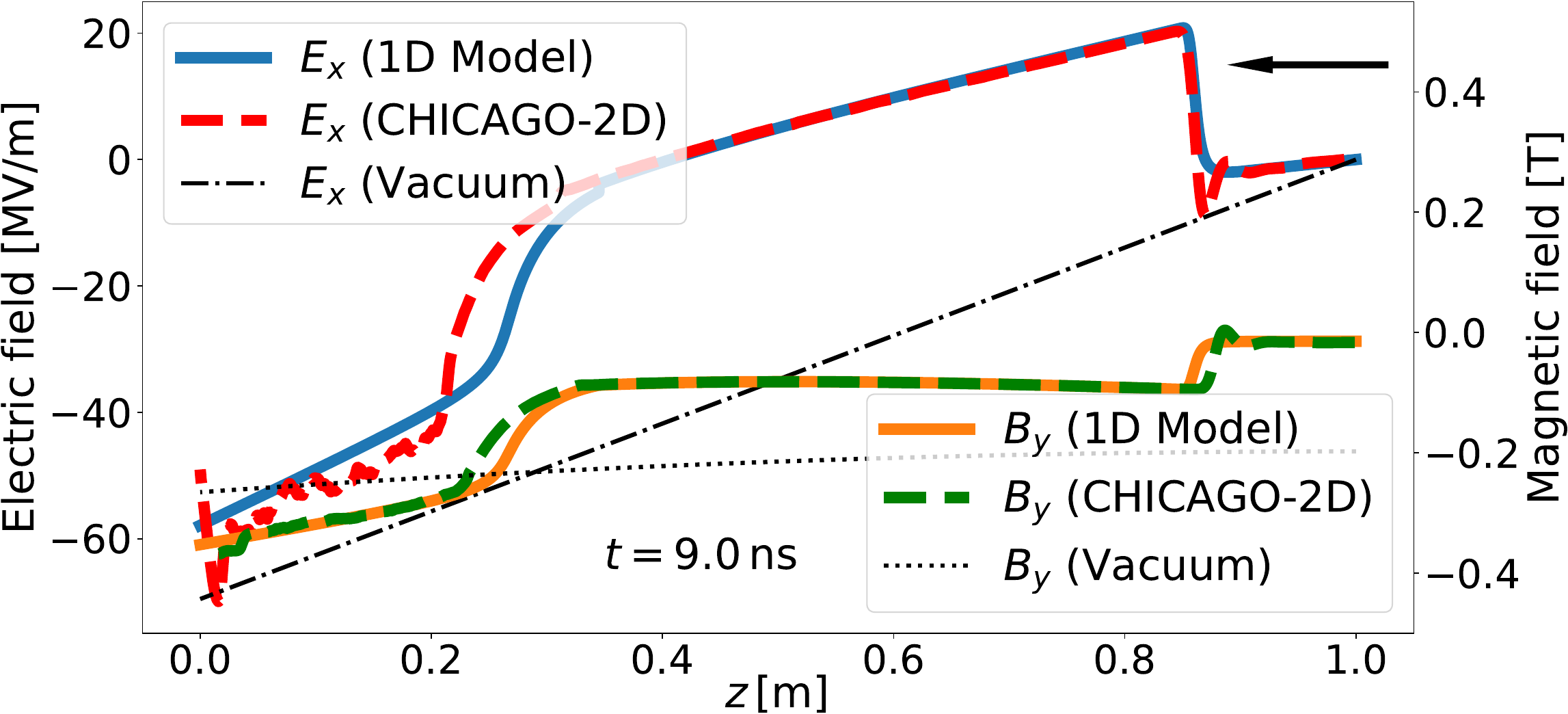}\vspace{5pt}\\
  \includegraphics[width=0.4\textwidth]{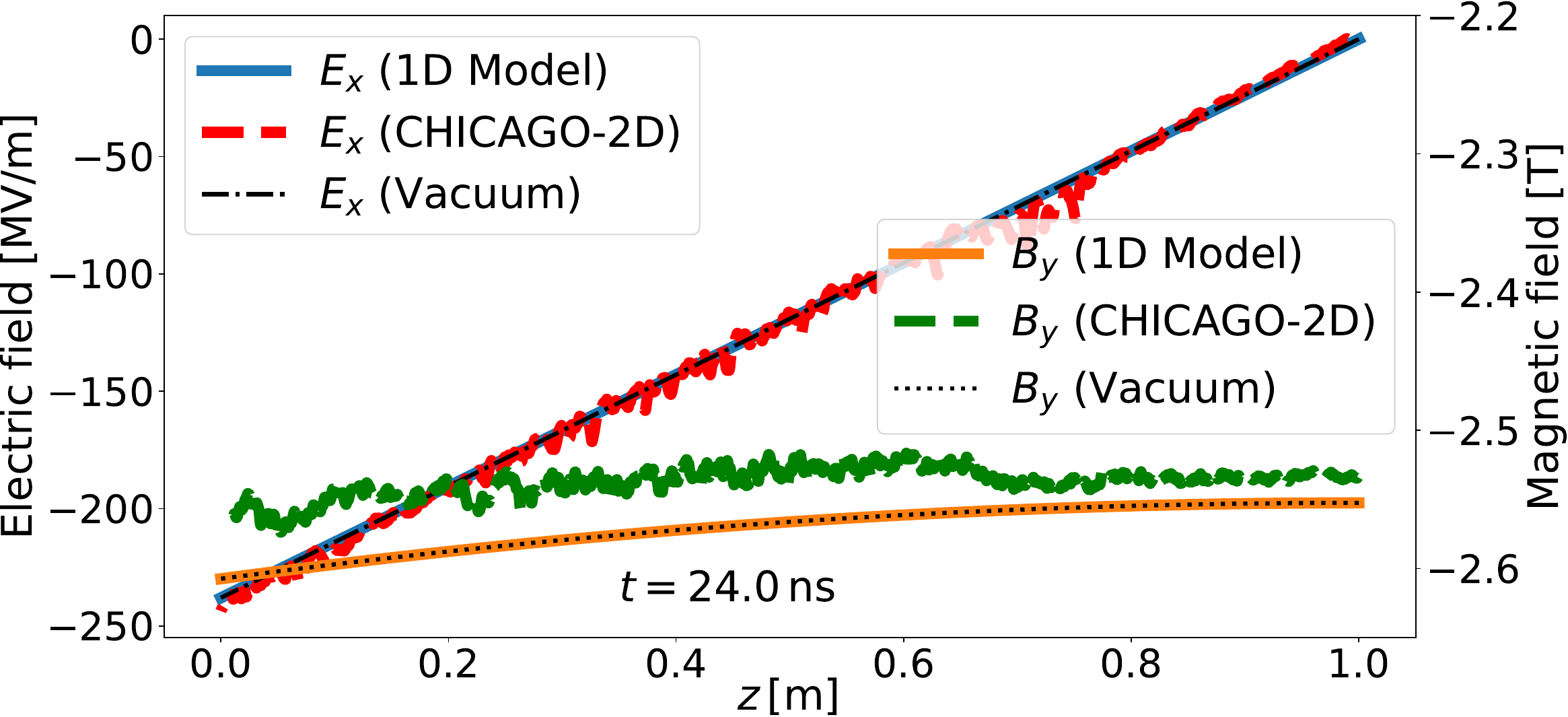}\\  
  \caption{Time evolution of the electric and magnetic fields in an SCL simulation. Top panel: the initial diamagnetic
    electromagnetic wave emitted by initial SCL currents, at $6.5\,$ns. Middle panel: The diamagnetic wave after
    reflection from the conducting load, at $9\,$ns. Bottom panel: electromagnetic fields after magnetic insulation
    approaching the vacuum (external) fields, at $24\,$ns.}
\label{fig:SCL_fields}
\end{figure}

The large diamagnetic response of the SCL currents is demonstrated in Fig.~\ref{fig:SCL_fields} at three times,
$6.5\,$ns, $9\,$ns, and $24\,$ns.
For this setup, we use a sine squared drive current pulse with $\Ipeak=20\,$MA/m and $\Taupeak=100\,$ns.
\tcb{SCL emission starts at about $5\,$ns. At $6.5\,$ns,}
the electric field measured near the anode drops to near zero, indicating
that the diamagnetic electric field amplitude is
comparable to the threshold for SCL emission  (top panel).
\tcg{(Unless specified otherwise, all simulations assume threshold for SCL emission $24\,$MV/m.)}
The middle panel of the figure shows the diamagnetic wave after reflection from the conducting short at $z=1\,$m.
The bottom panel shows the electromagnetic fields after magnetic insulation, which approach the
vacuum fields  (black lines).
\tcb{The bottom panel also shows that the electric field more closely approximates the vacuum
  field, while the magnetic field exhibits a small offset, differing by less than 1\% near $z=1\,$m to a maximum
  of about 1.6\% near $z=0\,$m. This offset can be explained by two
  observations. First, after magnetic insulation, the electrons form a layer near the cathode; that layer
  creates a magnetic field (also referred to as self-magnetic field in the literature),
  which our 1D model neglects. Second, numerically it is difficult to resolve
  the electromagnetic fields near the time of insulation (just before or shortly after)
  due to the fact that the statistics of computational particles reaching the anode
  becomes very poor, leading to increased numerical noise and subsequent accumulation of numerical errors that degrade
  the accuracy of the simulation.}

We may note the similarity between the diamagnetic fields in Figs.~\ref{fig:single_wave} and \ref{fig:SCL_fields},
the difference being that in Fig.~\ref{fig:SCL_fields}, the time-dependent external drive
causes SCL currents to \textit{accumulate} in time 
and to spread over a finite emitting cathode surface.

\begin{figure}[t!!]
  \centering
  \includegraphics[width=0.4\textwidth]{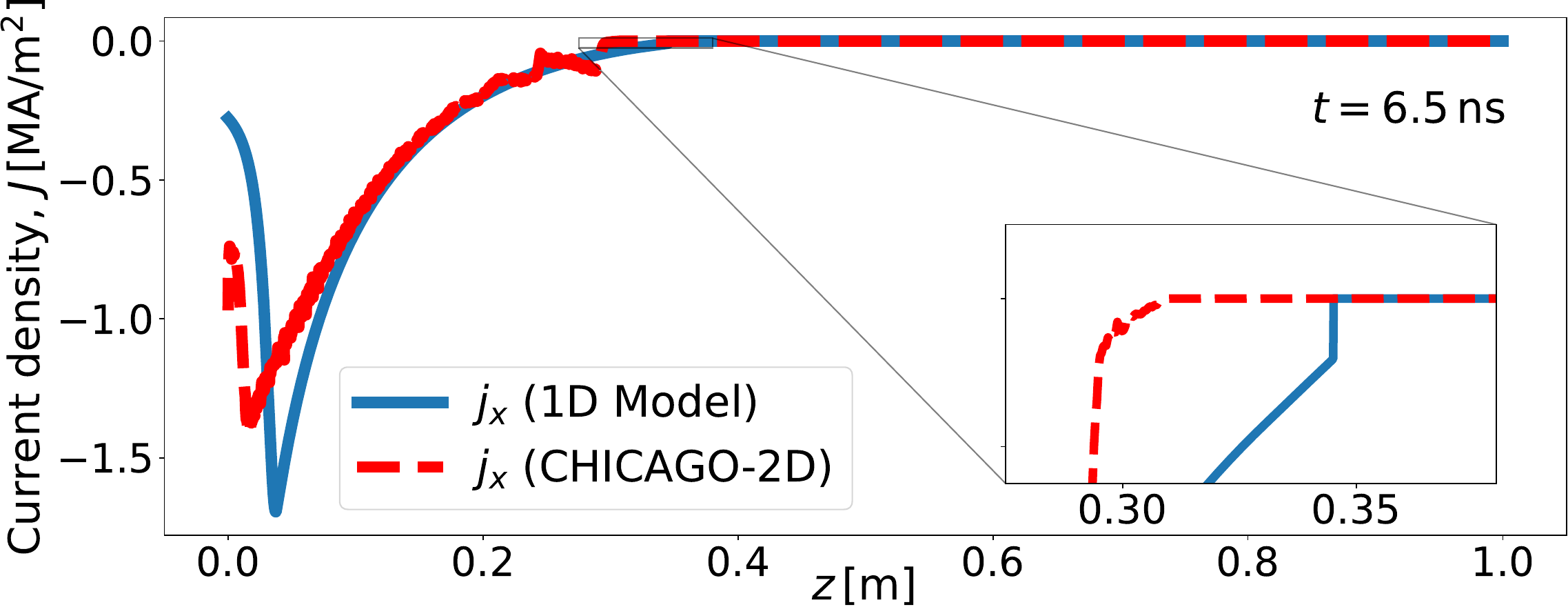}\vspace{5pt}\\
  \includegraphics[width=0.4\textwidth]{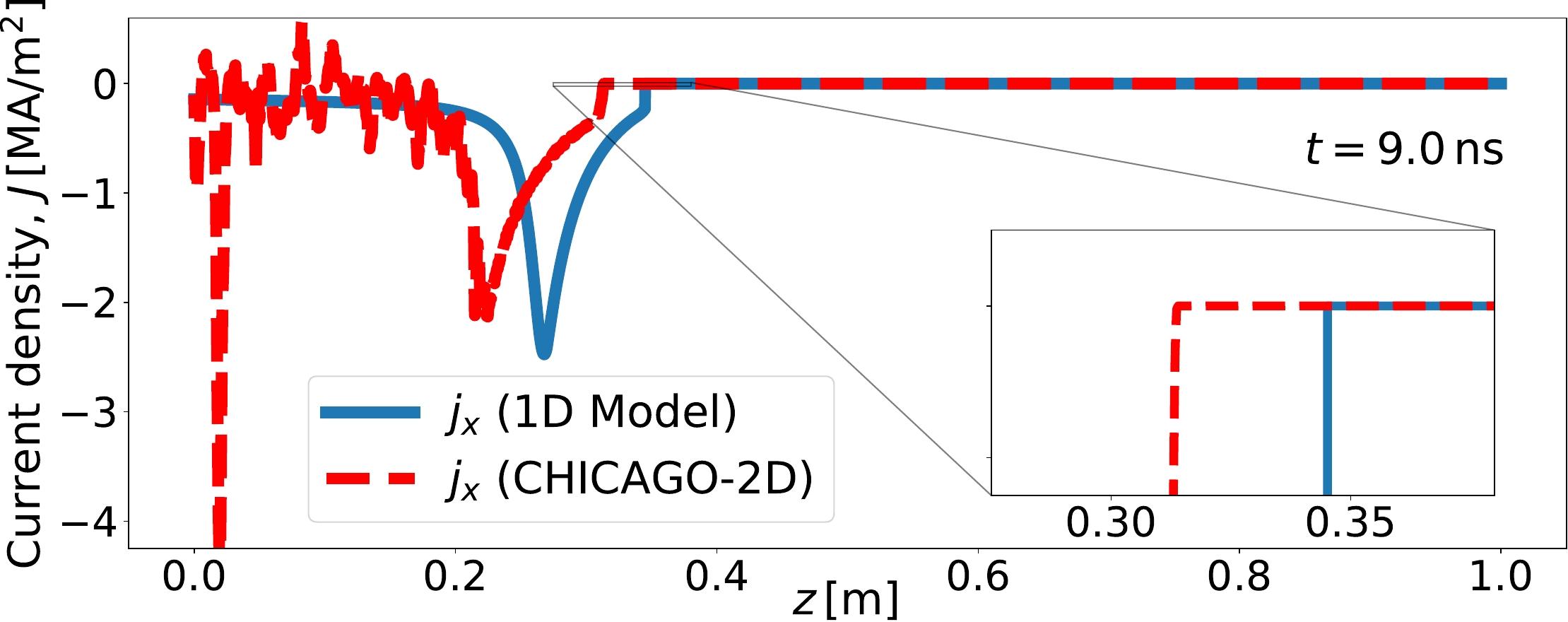}\\  
  \caption{Time evolution of the loss current density. Top panel: loss current density profile at $6.5\,$ns.
    Bottom panel: loss current density profile at $9\,$ns. It is seen that the peak of the loss current
    density propagates at about $0.65c$, however, the insets of the two panels show that SCL emission
    is inhibited past about $0.3\,$m in CHICAGO or $0.35\,$m in the 1D simulation. The latter phenomenon
    is quite common and is a result of the non-local diamagnetic EM effects by SCL currents.}       
\label{fig:SCL_current_densities}
\end{figure}

Another consequence of the diamagnetic effect is shown in Fig.~\ref{fig:SCL_current_densities}, where current
densities at the anode along the extent of the MITL are shown.
Non-zero SCL current densities are also an indication of where
the SCL emission threshold has been exceeded. The top panel shows that in about $1.5\,$ns
of (uninsulated) SCL emission (we remind the reader that SCL emission initiates at about $5\,$ns),
about $30\,$cm of cathode (surface) extent has started emitting
according to the PIC model and about $35\,$cm in the 1D model.
At time $9\,$ns, shown in the bottom panel, since the diamagnetic electric fields subtract from the vacuum fields,
the total resulting electric field in the remaining part of the MITL extent ($\gtrsim 30\,$cm)
has fallen below the emission threshold, halting 
SCL emission almost completely for the next $3\,$ns.
The vacuum fields at $9\,$ns are shown in the middle panel of Fig.~\eqref{fig:SCL_fields}
and suggest that without this diamagnetic effect, SCL emission would have covered
the extent of about $0$--$60\,$cm of cathode surface.

Such dynamics of fields and currents
tends to repeat itself, causing the emitting fraction of MITL surface to grow in ``jumps'' instead
of gradually, as one might expect from a smooth temporal drive.
In other words, for the time interval before the full cathode surface starts emitting
(this is roughly the time for the first diamagnetic wave to reach the load),
there is a sequence of a propagating SCL current pulse followed by a pause.
During the propagating part of the cycle, the diamagnetic effect accumulates until
the total electric field amplitude falls bellow the SCL emitting threshold.
During the pause part of the cycle, the external fields grow sufficiently so that
the total field overwhelms the diamagnetic effect and the fraction of emitting surface extends
further into the MITL.

We see that the non-local diamagnetic response acts to lower
current losses in two ways.
First, by limiting the extent of emitting surface.
This effect varies in magnitude in various setups, however,
it can be used as a telltale sign of the non-local, diamagnetic response of SCL currents.
And second, by lowering the electric field magnitude within the emitting surface,
leading to lower SCL currents (recall $j_{\rm SCL}\sim E^{3/2}$).
This latter effect is \textit{bi-directional} since local SCL currents
emit diamagnetic EM waves in both propagating directions. The result of this
is observed in Fig.~\ref{fig:SCL_current_densities} (bottom panel)
where the current density near the launch side ($z=0$) has been greatly reduced,
forming a sharp peak near $z\simeq 30\,$cm.
It is the combination of these two effects that causes dramatically lower
current losses in long MITLs compared to losses by SCL emission in vacuum fields alone
(see also the discussion of Fig.~\ref{fig:SCL_current_loss} below).

\tcb{The bi-directional diamagnetic property also affects the magnetic fields within the MITL,
  as observed in Fig.~\eqref{fig:SCL_fields}. The top and middle panels of the figure show
  the magnetic fields in the MITL with and without SCL emission.
  We observe in both panels magnetic field enhancement (in absolute value) near $z=0$,
  while the middle panel also shows magnetic field reduction near $z=1\,$m.
  This is understood by referring to Fig.~\ref{fig:diamagnetic} and the top and middle panels of
  Fig.~\ref{fig:single_wave}. Namely, we see that the magnetic field of an emitted to the left diamagnetic wave adds, 
  while that of a diamagnetic wave emitted to the right (including after reflection)
  subtracts from the magnetic fields of the external wave.}


Some interesting observations can relate our results to previous work.
First, the peak of the uninsulated current density propagates about $23\,$cm in $1.5\,$ns,
which is approximately at the speed of $0.65c$.
The current pulse observed (on the conductor) in the MITL is the difference between
the vacuum current and the loss current, and it would therefore appear that the main current pulse
propagates at that speed as well.
Second, since the the SCL (loss) current is seen to have a sharp peak, the main current pulse
would exhibit ``front sharpening.'' Both features have been previously reported
by experimental and numerical work
\cite{baranchikov_transfer_1977,di_capua_propagation_1979,bergeron_equivalent_1977,luo_properties_2017}.
On the other hand, the ``jump-like'' pulse propagation discussed above has not been previously reported.

\section{Scaling studies of {MITL} losses based on the 1D electromagnetic model}
\label{sec:scalings}

We have tested relatively extensively the validity of the 1D model against PIC simulations,
varying parameters such as MITL dimensions (length, AK gap), peak current, 
pulse length, as well as threshold for SCL emission within relevant ranges.
In \tcg{tests with} very short MITLs, for example, lowering $\Eth$ to artificially low values was necessary
in order to observe uninsulated SCL emission. 

\begin{figure}[t!!]
  \centering
  \includegraphics[width=0.4\textwidth]{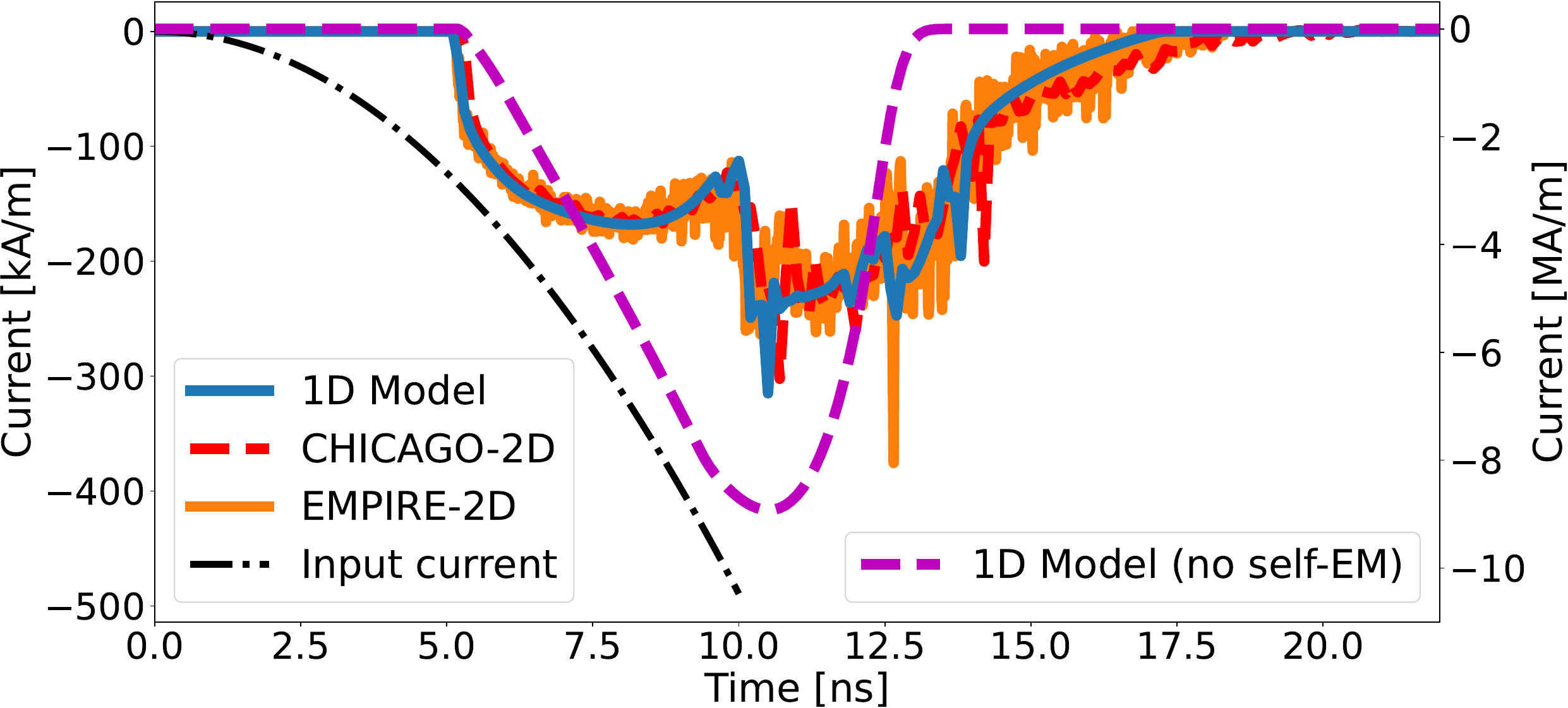}
  \caption{Current loss vs. time: Comparison between the 1D model and 2D PIC simulations with EMPIRE and CHICAGO
    (left-hand scale). The dashed magenta curve (right-hand scale) shows the current loss without accounting for
    non-local diamagnetic effects.}
\label{fig:SCL_current_loss}
\end{figure}

We define current loss as 
\begin{equation}
  I_{\rm loss}(t) =  \int_0^L\!\! dz\, j_{\rm SCL}(z,t)
  \label{eq:I_loss}
\end{equation}
where $j_{\rm SCL}(z,t)$ is the \textit{uninsulated} current 
and note that it is a function of time alone. A comparison of predicted current loss
by the 1D model as well as EMPIRE and CHICAGO 2D PIC is shown in Fig.~\ref{fig:SCL_current_loss}
(left-hand scale). We note the excellent agreement between all three codes. The input current pulse
(without SCL emission) is indicated with a dash-dotted black line.

Also plotted in Fig.~\ref{fig:SCL_current_loss} is the loss current (by the 1D model)
based on only space charge limiting effects, i.e., the Child-Langmuir law (right-hand scale).
We see that such a prediction, even for a $1\,$m long MITL, is grossly inaccurate, with a
maximum current loss of over $8\,$MA/m, compared to the expected $250\,$kA/m.
\tcb{That the loss current exceeds the input current should not be a cause for alarm.
  Such a setup is obviously not physical for the parameters at hand 
  and in an electromagnetic simulation the loss current never exceeds the drive current. 
  However, one may think of this as a quasistatic approximation, where the input drive changes
  so slowly that electromagnetic effects can be neglected; alternatively, one may
  think of this approximation as the limit $c\rightarrow \infty$. In this limit, the input
  drive acts as an \textit{ideal voltage source}, which can provide arbitrarily large current
  to a circuit. Although not valid for the present case, the quasistatic approximation
  becomes valid in the limit of an infinitely short MITL. Indeed, we have verified that
  in this limit the quasistatic and electromagnetic simulations yield similar current
  losses (not shown).}

In this section we look at some scaling laws that could suggest appropriate MITL design parameters
and provide additional insight into the time-dependent effects in long MITLs.
We would like to compare a quantity that characterizes MITL losses in both space and time.
For this purpose, we find it convenient to define the \textit{loss charge} as
\begin{equation}
  Q_{\rm loss} =  \int_0^{t_{\rm ins}}\!\! dt \, I_{\rm loss}(t).
  \label{eq:q_loss}
\end{equation}
The integral's upper limit is the time 
\tcb{for complete magnetic insulation along the MITL, $t_{\rm ins}$, defined as 
$I_{\rm loss}(t) = 0$ for $t>t_{\rm ins}$};
then $Q_{\rm loss}$ is the total charge collected by the anode.
We normalize this to the total charge delivered by the drive current, which
for a sine squared pulse equals $Q_{\rm tot} = \Ipeak \Taupeak$.

\begin{figure}[t!!]
  \centering
  \includegraphics[width=0.4\textwidth]{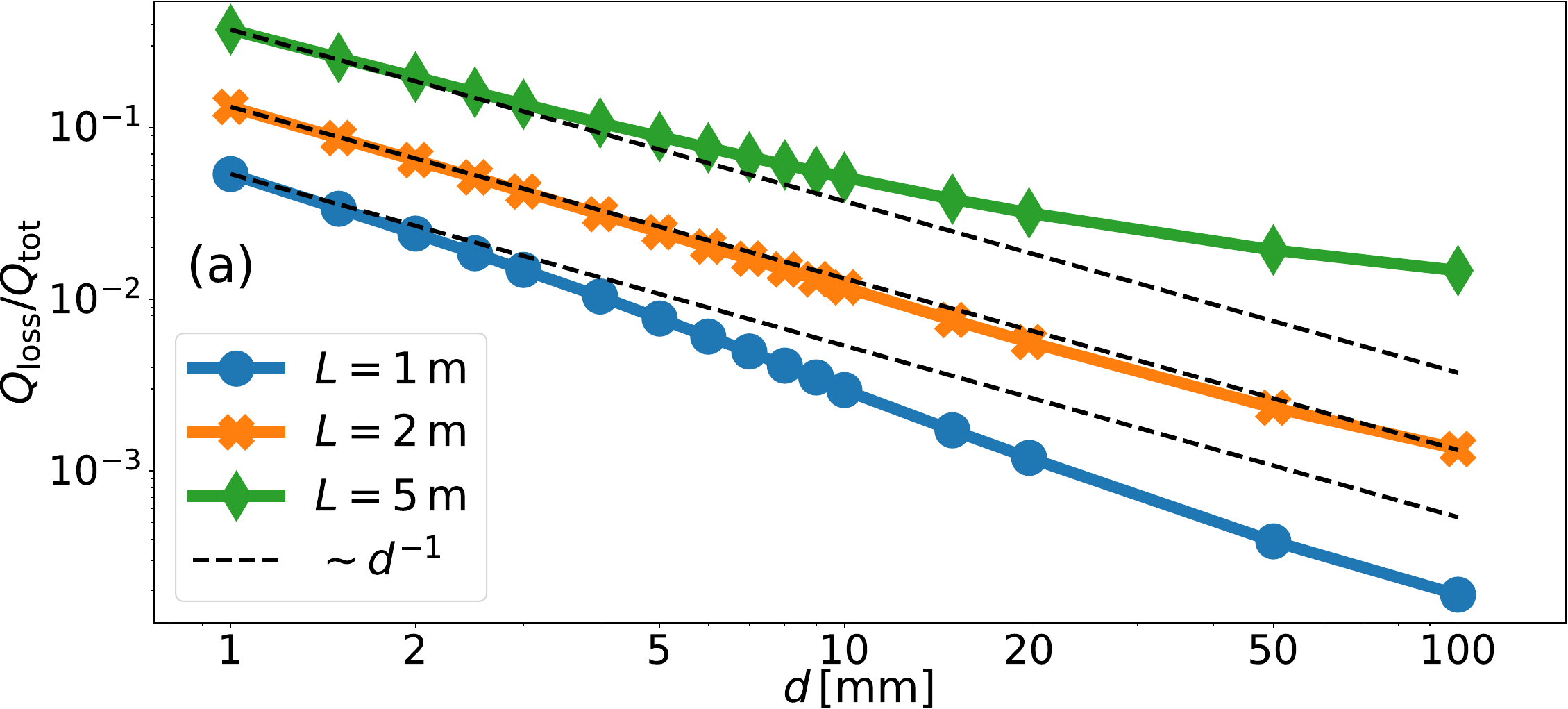}\vspace{5pt}\\
  \includegraphics[width=0.4\textwidth]{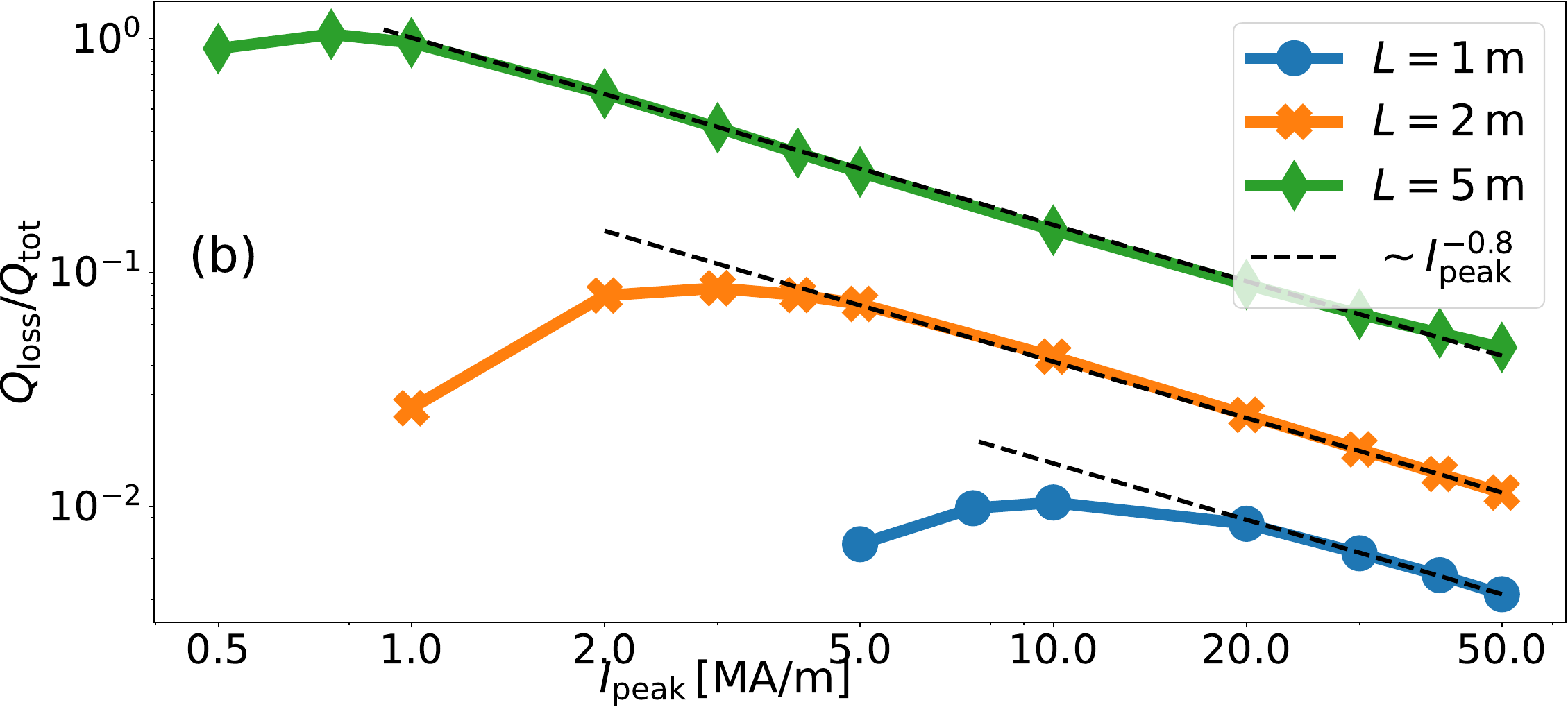}\vspace{5pt}\\
  \includegraphics[width=0.4\textwidth]{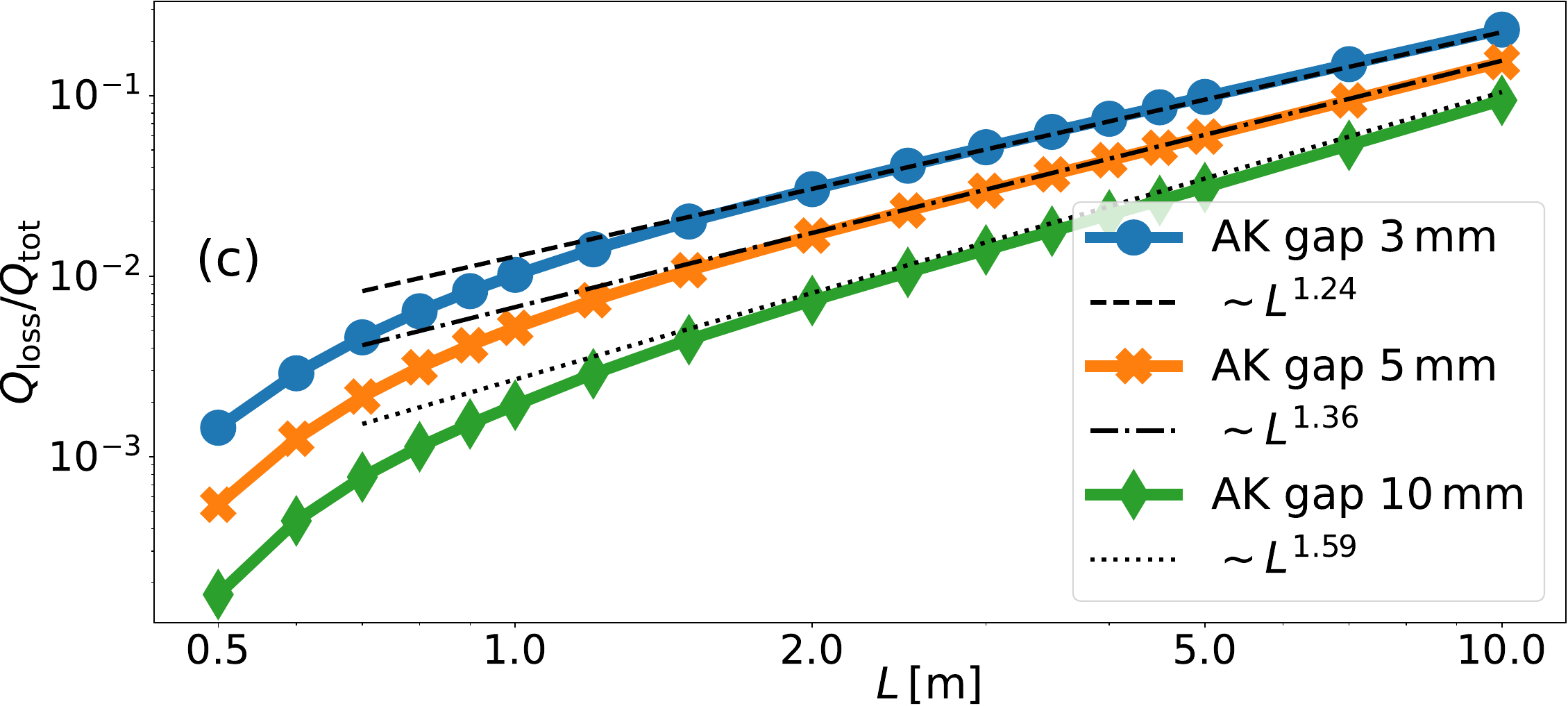}\vspace{5pt}\\
  \includegraphics[width=0.4\textwidth]{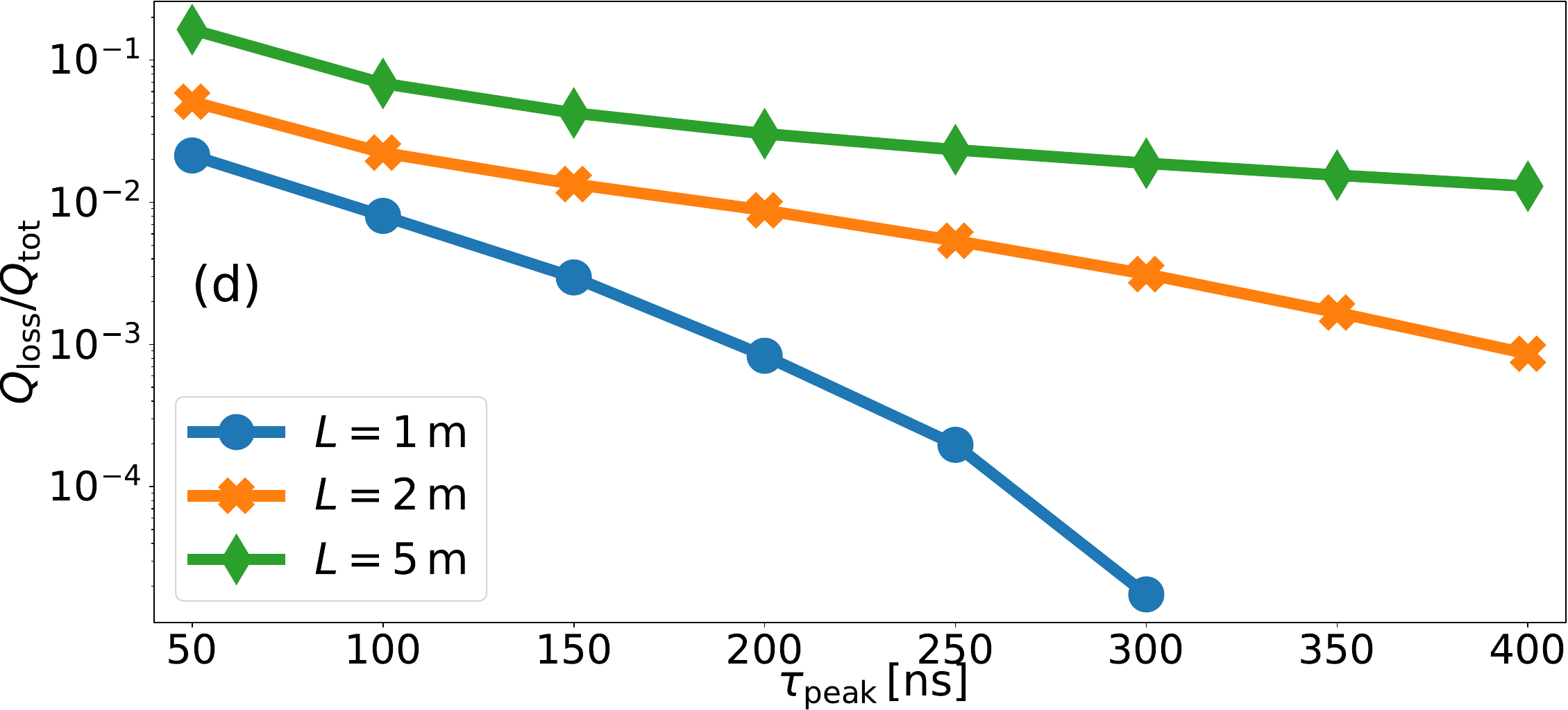}
  \caption{Scaling studies of loss charge. Panel (a): Scaling with AK gap, $d$,
    for fixed $\Ipeak=20\,$MA/m and $\Taupeak = 100\,$ns. Panel (b): Scaling with peak current, $\Ipeak$,
    for a fixed AK gap, $d=5\,$mm and $\Taupeak=100\,$ns. Panel (c): Scaling with pulse length,
    $\Taupeak$, for a fixed $\Ipeak=20\,$MA/m and AK gap, $d=5\,$mm.
    Panel (d): Scaling with MITL length, $L$, for a fixed $\Ipeak=20\,$MA/m
    and $\Taupeak = 100\,$ns.
    }
  \label{fig:scalings}
\end{figure}

In Fig.~\ref{fig:scalings} (a) we show the scaling of MITL losses with the AK gap distance, $d$, for
three different lengths. The pulse has $\Ipeak=20\,$MA/m and $\Taupeak=100\,$ns. We observe
that for small AK gap sizes, $Q_{\rm loss}/Q_{\rm tot}$ scales as a power law $\sim 1/d$ while for larger gap sizes
that scaling does not hold. In practical applications, small gap sizes
are usually preferred and then the \tcb{$\sim 1/d$} scaling can be used.

Fig.~\ref{fig:scalings} (b) shows how MITL losses scale with peak current, $\Ipeak$,
for three (fixed) MITL lengths, pulse length $\Taupeak=100\,$ns, and $d=5\,$mm.
The fractional losses exhibit a maximum at a certain value of $\Ipeak$, vanish
for small currents, and decrease for large peak currents. We can understand this
behavior in the following way. On the one hand,
smaller peak currents have smaller losses due to the shorter time of uninsulated SCL emission:
regardless of the drive current amplitude, SCL emission turns on only after the electric field
exceeds some fixed threshold value. 
If the peak current is too low, that threshold value cannot be exceeded
since for a fixed $\Taupeak$, $E \sim \Ipeak$.
On the other hand, large peak currents induce larger diamagnetic
effects, which lower the losses. Obviously, peak currents around the maximum
of the curve should be avoided in MITL designs.
A common scaling of MITL losses for large peak currents emerges in the form of \tcb{$\sim \Ipeak^{-0.8}$},
which we leave as an empirical observation.

In Fig.~\ref{fig:scalings} (c) we show the scaling of MITL losses with MITL length for fixed
$\Ipeak=20\,$MA/m and $\Taupeak=100\,$ns, for three different AK gap sizes.
At shorter lengths, the current losses drop due to the decreased time of uninsulated
SCL emission. In the limit of (infinitely) short MITLs, we converge
to the losses predicted by the Child-Langmuir law in vacuum fields
(not shown; see also the discussion of Fig.~\ref{fig:SCL_current_loss}).
Similarly, we can understand the increased current losses in longer MITLs as increased
time of uninsulated SCL emission. Recall that magnetic insulation cannot happen in pure
TEM fields, i.e., for which $E=cB$: in such fields, electrons perform a figure-8 motion \tcb{\cite{landau_lifshitz_vol2}},
not a cyclotron motion. Our estimates have shown that the size of such a figure-8 orbit for typical
fields of interest is much larger than the AK gap, i.e., such electrons are uninsulated.
Magnetic insulation can occur \textit{after reflection} from the load because
the relation $E=cB$ no longer holds due to
the interference (superposition) of the incoming and outgoing (reflected) waves.
In longer MITLs, reflection takes longer to happen, hence, longer time of uninsulation and increased losses.
For the range of lengths shown, we again uncover an empirical scaling law for large $L$, however,
different for the different gap sizes.

It is worth commenting on the losses in \textit{infinitely long} MITLs.
Although magnetic insulation never occurs in such MITLs,
we reiterate that some of the current limiting mechanisms already discussed in
sections \ref{sec:1D_model} and \ref{sec:diamagnetic} still apply:
(i) the Child-Langmuir law limits the amount of current grossing the AK gap;
and (ii) the non-local diamagnetic effects further lower the amount of loss current;
in fact, the latter are maximally manifested in this case.

Figure \ref{fig:scalings} (d) shows scaling of MITL losses with pulse duration.
This scaling is the strongest of the four examples shown in the figure, showing exponential
or super-exponential dependence (notice the semi-log scale of the plot).
The losses for the $1\,$m-long MITL fall to zero past about $\Taupeak\gtrsim 300\,$ns.
These losses can also be understood in terms of uninsulation time.
First, uninsulation time is longer in longer MITLs (as discussed above),
hence, increasing losses for larger $L$.
Second, the relation $E\sim dI/dt$ implies lower fields for longer pulse lengths
(for a fixed AK gap), hence, shorter uninsulation times for longer pulse lengths.
(Strictly speaking, this relation only applies in infinitely short MITLs or equivalently,
in the limit of $c\rightarrow \infty$, see also Ref.~\cite{hess_electron_2021}.
However, it approximately applies for the temporal spatial MITL lengths considered here,
while also giving an intuitive way of discussing the observed dependencies.)
No common empirical scaling law emerges in this case.

\section{Magnetic insulation model}
\label{sec:hull_curve}

The magnetic insulation model is a common ingredient in several major circuit element models.
Since it is also used in our 1D model, it deserves special attention.

The issue at hand is the value of the SCL current crossing the AK gap
as a function of magnetic field magnitude.
A theoretical value of the critical magnetic field
(also called the Hull field or Hull cutoff \cite{hull_magnetic_insulation_1921}),
at which complete insulation occurs, was derived by Lovelace and Ott \cite{lovelace_theory_1974}.
\tcg{(An earlier calculation, based on a single particle motion in external fields,
  was given by Walker \cite{collins_magnetrons_1948}.)}
The expression they obtained depends on the value of the potential (difference), $V_0$,
[cf. Eqs.~\eqref{eq:j_CL} and \eqref{eq:V_CL}] and is given by
\begin{equation}
  B_{\rm crit} = \frac{mc}{ed}\sqrt{\left(\frac{eV_0}{mc^2}\right)^2 + \frac{2eV_0}{mc^2}}.
  \label{eq:Lovelace_Ott}
\end{equation}
This dependence was experimentally tested by Orzechowski
and Bekefi \cite{orzechowski_current_1976} who found that magnetic insulation
occurred mostly in agreement with Eq.~\eqref{eq:Lovelace_Ott} but that the SCL current did not
sharply drop to zero at the Hull magnetic field value; instead, it had a ``spillover''
with non-zero current reaching the anode past the critical cutoff.
Reference \cite{orzechowski_current_1976} presents curves of
SCL current vs. magnetic field in normalized units, whereby
their measured current was normalized to either $V_0^{3/2}$ (termed \textit{perveance})
or to the Child-Langmuir current; they normalized the measured magnetic field to the
theoretical critical magnetic field\eqref{eq:Lovelace_Ott}.
In the following, we normalize our SCL currents to the Child-Langmuir current
and magnetic fields to $B_{\rm crit}$.
We refer to the curve of normalized emitted SCL current vs. normalized magnetic
field as the Hull curve.
Since the expression for the Child-Langmuir current folds in the particular geometry,
one can think of such a curve as universal. 

\begin{figure}[t!!]
  \centering
  \includegraphics[width=0.4\textwidth]{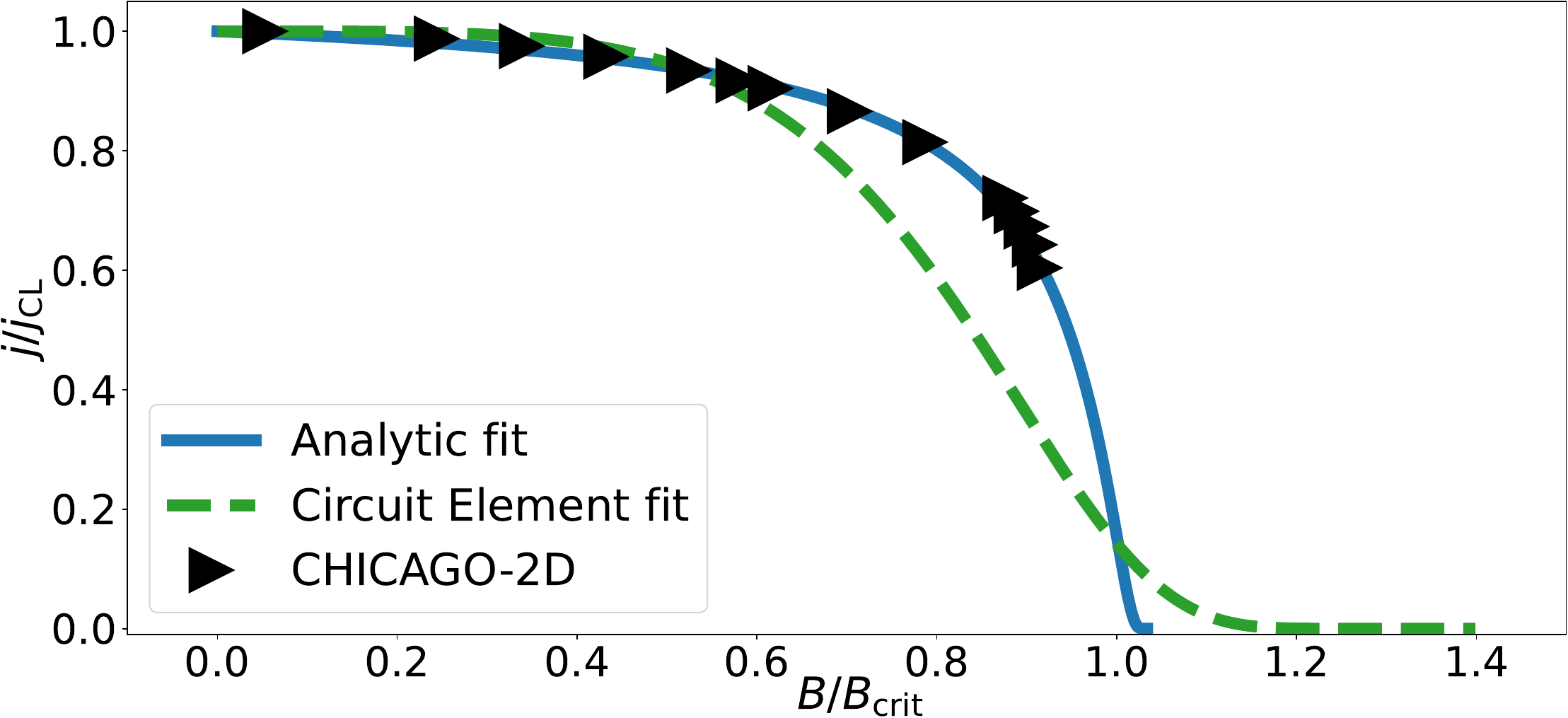}
  \caption{Comparison of the Hull curve used in our 1D model, 'Analytic fit', and
    the Hull curve used presently in circuit element codes, 'Circuit Element fit'.
    CHICAGO 2D PIC simulations are shown with black triangles.}
\label{fig:hull_curve}
\end{figure}

As CEM is concerned in regards to 
predicting SCL current losses in MITLs, we note that major codes, such as BERTHA
and SCREAMER, use the curve plotted in Fig.~7 of
Ref.~\cite{vandevender_requirements_2015}, labeled ``Original SCREAMER.''
Reference \cite{vandevender_requirements_2015} also proposes a revised version of the curve,
however, such revision does not appear to have been widely implemented \cite{spielman_private_2024}. 
We refer to the curve ``Original SCREAMER'' from Ref.~\cite{vandevender_requirements_2015}
more generally as the ``Circuit Element fit.'' This curve is shown as a dashed
line\footnote{Our fit of that curve is based on
  the image from Ref.~\cite{vandevender_requirements_2015}.\label{footnote:Hull_curve}}
in Fig.~\ref{fig:hull_curve}.
Reference \cite{vandevender_requirements_2015} additionally states that the Hull curve
has been constructed from 3D PIC simulations. We attempted to replicate
the curve using 2D PIC simulations with CHICAGO and found that
our results did not agree with the Circuit Element fit curve,
as indicated by the black triangles in Fig.~\ref{fig:hull_curve}.

One aspect of the disagreement is worth discussing. One can see that our simulation data
stops at about $B/B_{\rm crit}=0.91$. The reason behind this is that as we approached the
Hull magnetic field, the electron layer transitioned from stable and steady-state to unstable.
In an unstable layer, the current collected by the anode as well as the electric field
acquire oscillatory behavior. (The magnetic field is much more constant since it is externally
imposed and typically much larger than the (self-)magnetic field of the SCL currents.
There is no contradiction with the results in the previous sections since
here the electromagnetic effects were negligible due to the system being
in either a steady state or having slow fluctuations when unstable. In addition, the system length
in these simulations was quite short, about $10\,$cm, which also necessitated using an artificially low
threshold value of $380\,$kV/m.)
In such an unstable system, averaged values are difficult to obtain due to the
randomness in the oscillations. Even if averaged values were possible to reliably obtain,
they would not be representative of a steady-state system, upon which assumption
both the Child-Langmuir law and Hull magnetic field calculations are based.
However, knowing an accurate value of $V_0$ is crucially important
since both the Child-Langmuir current and the critical magnetic field
strongly depend on it, as seen from Eqs.\eqref{eq:j_CL} and \eqref{eq:Lovelace_Ott}.
These two quantities are being used to normalize the measured current and magnetic field,
and because of that, any uncertainty in $V_0$ translates into uncertainty in the normalized
quantities on the Hull curve. Our experience has shown that near cutoff,
even a small uncertainty in $V_0$ can easily place a point on either side of the theoretical Hull field.

Apart from the uncertainty associated with the value of $V_0$, our PIC simulations
\footnote{\tcb{Our simulations with CHICAGO were electromagnetic and a wave was launched into the
  simulation domain using a simple input circuit.
  The value of the electric potential $V_0$ entering Eqs.~\eqref{eq:j_CL}, \eqref{eq:V_CL},
  \eqref{eq:Lovelace_Ott} was calculated using relation \eqref{eq:V_0} with  $\widetilde{E}_x(x=d;z,t)$
  provided by CHICAGO.}}
confirm the ``spillover'' effect and the loss of a sharp insulating boundary.
These observations have led us to revisit the \textit{interpretation} of the experimental results in
Ref.~\cite{orzechowski_current_1976}. In that reference, it was hypothesized that
some process, not characterized by the authors, produced ``hot tail'' electrons that
contributed to smearing of the sharp theoretical boundary.
The authors detected radiation
when the magnetic field strength exceeded the Hull field and
no such radiation for lower magnetic fields,
which they also correlated with the onset of an instability in the system.
To our knowledge, a precise mechanism for such ``hot electrons''
has not been identified in the literature to date.
Here we suggest another possible mechanism leading to the loss of a sharp
insulating boundary. Namely, we propose that the electron layer loses stability due to a
velocity shear type instability in crossed field devices, such as diocotron, magnetron, etc.
In such an unstable layer, our observations show the formation of large-scale
vortical structures that move in a stochastic way throughout the system (largely because of
$\mathbf{E}\times\mathbf{B}$ drift), leading to stochastic
oscillatory behavior in all quantities within the AK gap,
including the conducting current.
Such mechanism also results in non-zero loss current beyond the Hull field but
does not necessarily produce a population of hot electrons.
Ultimately, to be able to unambiguously identify the precise instability,
a further study beyond the scope of this work is necessary.

The stability properties of a parapotential magnetized electron layer have been discussed in the
past \cite{buneman_stability_1966,davidson_equilibrium_1991,davidson_physics_2001}.
While these properties depend on the layer's spatial profile, of equal importance is
whether and how this profile may in practice (experiment) be attained.
Many factors make this problem very complex, however, for relevant to $Z$ parameters,
even the simplest constant profile electron layer is never strictly stable,
something recently emphasized in Ref.~\cite{evstatiev_efficient_2023}.
(See also reference \cite{revolinsky_two_dimensional_2024}, which considers vortex formation in
geometries with an abrupt change in the AK gap size.)
An attempt to create a stable electron layer
by the authors of Ref.~\cite{bergeron_beam_1979} was also not successful.
\tcg{Strong indication of a possible change in the stability properties
  of the electron layer in the AK gap can be seen in the analyses of the Hull curve
  \cite{christenson_transition_1994,lau_limiting_1993,lau_brillouin_2021};
  for example, a jump in the SCL current around the value of near $B_{\rm crit}$  was noted in
  Ref.~\cite{christenson_transition_1994}, while Refs.~\cite{lau_limiting_1993,lau_brillouin_2021}
  note the infinite slope of the curve at $B_{\rm crit}$. An earlier work by
  Pollack and Whinnery \cite{pollack_noise_1964} observed both experimentally and computationally
  a sharp increase in the noise level in a planar diode as an externally imposed
  magnetic field exceeded the value of the Hull cutoff.}
We believe that in order to properly understand the loss of a sharp insulating boundary,
a theory of current transport across an AK gap in unsteady flows must be developed.

Let us discuss the choice and analytic representation of our Hull curve.
Our fit is based on the formula
\begin{equation}
  Y(X) = \left\{
    \begin{tabular}{ll}
      $e^{-a |X|^\alpha/\left|X-b\right|}$, & $X\le 0 <b$ \\
      $0$, & $X\ge b$
    \end{tabular}
              \right.
  \label{eq:Hull_fit}
\end{equation}
with positive $a$, $b$, and $\alpha$, and $X = B/B_{\rm crit}$ (in the Cartesian geometry,
the magnetic field strength $B \equiv \left| B_y\right|$).
Our Analytic fit Hull curve is obtained for values $a=0.065$, $b\in[1.03,1.04]$, and
$\alpha=1$; the Circuit Element fit is obtained for $a=0.78$, $b=1.4$, and $\alpha=4$.
It is seen that the function has a value of unity for zero magnetic field strength $B=0$.
The notable feature here is that our analytic fit has a much smaller range outside of the
critical magnetic field, exceeding it by only about $3$--$4$\%, compared to about $40$\% for the
Circuit Element fit.

\begin{figure}[t!!]
  \centering
  \includegraphics[width=0.4\textwidth]{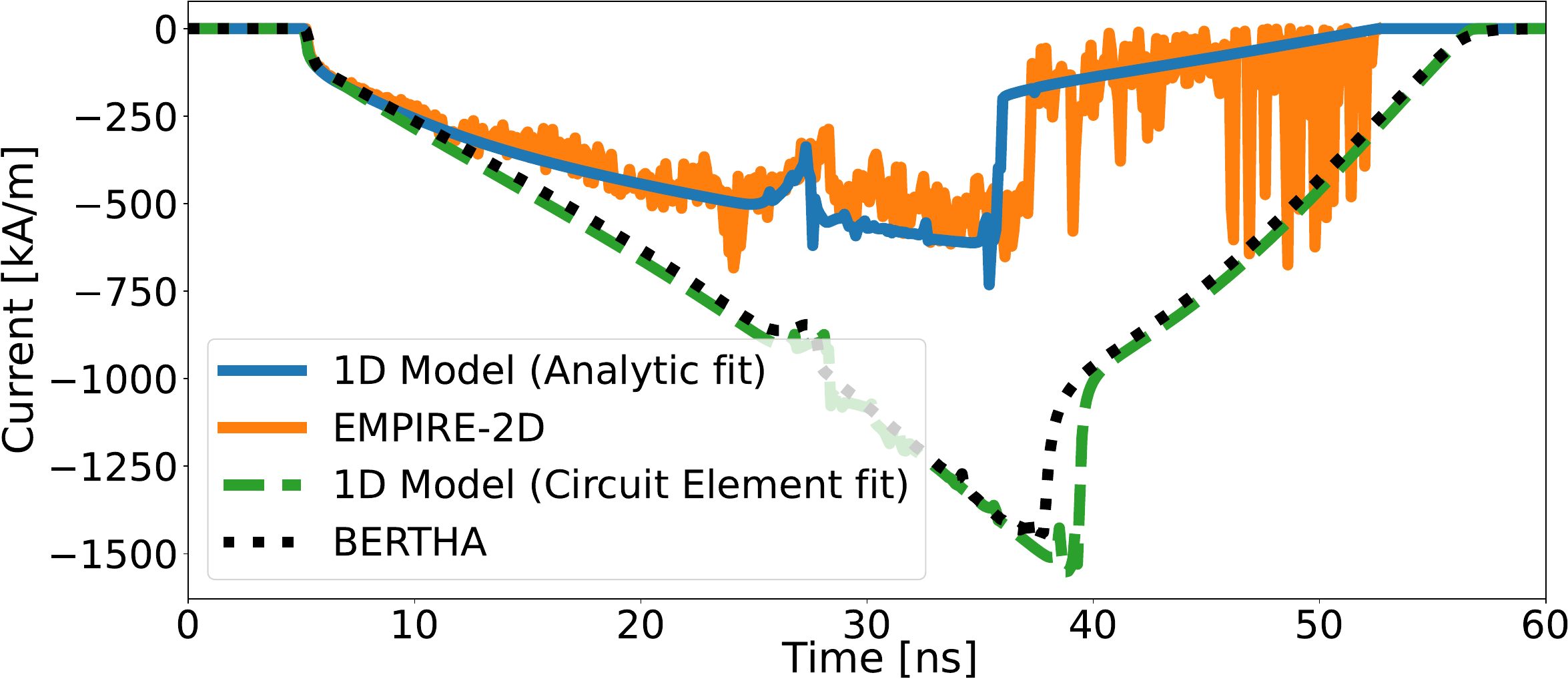}
  \caption{Comparison of current losses in a $5\,$m long MITL with
    $\Ipeak=20\,$MA/m, $\Taupeak=100\,$ns, AK gap, $d=5\,$mm. The Circuit Element fit
    overpredicts EMPIRE 2D PIC result by about a factor of $3$ while the
  Analytic fit is in excellent agreement with it.}
\label{fig:5m_MITL}
\end{figure}

Since our PIC simulations were not successful in constructing the entire Hull curve,
we have fitted the parameters in Eq.~\eqref{eq:Hull_fit} from PIC simulations
of current loss in MITLs of various geometrical dimensions.
In particular, long MITLs have an amplifying effect
on small variations in the curve parameter values.
Fig.~\ref{fig:5m_MITL} shows a simulation of current loss in a $5\,$m long MITL
($\Ipeak=20\,$MA/m, $\Taupeak=100\,$ns, AK gap, $d=5\,$mm), which 
illustrates the difference in the predictions using the two different fits
from Fig.~\ref{fig:hull_curve}. Clearly, the Circuit Element fit greatly overpredicts current
losses, by a factor of about $3$. (The small difference between the 1D model using the
Circuit Element fit and BERTHA is attributed to the fact that our model uses a fit, not the
actual Hull curve used in BERTHA, see footnote \ref{footnote:Hull_curve}.)
Conversely, the details of the Hull curve in shorter MITLs in the length range $\lesssim 2\,$m do not
manifest significant discrepancy with PIC simulations and both fits produce acceptable predictions.

One last remark in this section is that the ``spillover'' of
a Hull curve acts as a low pass filter, and the larger the ``spillover,''
the stronger the low pass filter effect.
For example, in Fig~\ref{fig:5m_MITL} we notice that the 1D model using the Analytic fit exhibits
certain small scale temporal oscillations (see also Fig.~\ref{fig:SCL_current_loss}). Such oscillations are
not PIC noise since the 1D model is a (cold) fluid model. \tcb{In the same figure,}
comparing with the simulations using
the Circuit Element fit, we see a noticeably smoother curve. We recall that
the Analytic fit exceeds $B_{\rm crit}$ by only about $3$-$4$\% while the
Circuit Element fit by about $40$\%.
This is not to imply that removing the small scale oscillations
is nonphysical, rather, that it is possible that certain physical oscillations may
unintentionally be filtered out.

\section{Temperature rise due to electron impact}
\label{sec:temperature}

Temperature rise due to electron impact can be calculated within our 1D model
as a diagnostic. Electrons impacting the anode deliver a certain amount of energy to
the surface, contributing to temperature increase.
\tcg{For the time scale of our current pulses,
a negligible amount of heat is dissipated within the metal volume, therefore,
our calculation simply accumulates the electron energy and temperature
impacted to the anode surface.}

We use the NIST stopping power tables for electrons in metals, $dK/dx$ \cite{nist_estar};
or for certain metals, we prefer to use an analytic expression that gives the extra
information in the low energy range \cite{nguyen-truong_modified_2015}.
First, the electron flux to the anode is directly related to the electron current:
\begin{equation}
  F(z,t) = \left|\frac{j_x(z,t)}{e}\right|.
  \label{eq:e_flux}
\end{equation}
The temperature increase due to electron impact in time $\Delta t$  at time $t_k$ is then
\begin{equation}
  \Delta T_k \!=\! \Delta T(z,t_k)\! =\!
  \frac{M_{\rm mol}\left|dK/dx\right| \int_{t_k}^{t_k+\Delta t}F(z,t)dt}{\rho C_V}\frac{1}{\cos \theta_i},
  \label{eq:delta_T}
\end{equation}
where $M_{\rm mol}$ is the molar mass of the metal, $\rho$ is the mass density,
$C_V$ is the heat capacitance at constant volume, and $\theta_i$ is the electron
angle of incidence measured from the normal to the surface
(see also Ref.~\cite{bennett_current_2019}). The total temperature increase
is the sum over all time steps, $N_{\rm ins} = t_{\rm ins}/\Delta t$, until complete
magnetic insulation is achieved,
\begin{equation}
  \Delta T^{\rm tot}(z) = \sum_{k=0}^{N_{\rm ins}} \Delta T_k(z,t_k).
  \label{eq:temperature}
\end{equation}

The only extra information necessary to calculate the temperature rise is the
angle of incidence. It can be calculated as follows. Using the steady-state
relations for conservation of the $z$-component of canonical momentum and the energy balance
in Cartesian geometry \cite{lovelace_theory_1974,bergeron_relativistic_1975}, we have
\begin{align}
  mv_z(x)\gamma(x) - eA_z(x) &= \mbox{const.} = 0, \label{eq:pz_conservation}\\
  mc^2\left[\gamma(x)-1\right] - eV(x) &= 0, \label{eq:KE_balance}
\end{align}
where the relativistic factor is defined as usual, $\gamma = \left[1-(v_x^2+v_z^2)/c^2\right]^{-1/2}$,
with electron velocities $v_x$ and $v_z$.
We note that the vector potential in Eq.~\eqref{eq:pz_conservation} is \textit{not} the same as
in Eq.~\eqref{eq:wave_eq}. For a constant magnetic field in the $y$-direction,
we can choose $\mathbf{A} = \left(0,0,-B_0x\right)$,
\footnote{One can easily extend this condition to spatially uniform, time-dependent electric fields by
  choosing $\mathbf{A} = \left(A_x(t),0,-B_0x\right)$, giving $E_x(t)=-dA_x(t)/dt$.}
which gives $B_y = \left(\nabla\times\mathbf{A}\right)_y = B_0$. We can calculate $\cos\theta_i$ as
\begin{equation}
  \cos\theta_i = \frac{v_x}{\sqrt{v_x^2+v_z^2}}.
  \label{eq:cos_i_def}
\end{equation}
Hereafter we calculate all quantities at the anode, $x=d$.
It is straightforward to find from \eqref{eq:pz_conservation} and the definition of $\gamma$
\begin{equation}
  v_z = \frac{e}{m}\frac{B_y d}{\gamma}, \qquad \sqrt{v_x^2 + v_z^2} = \frac{c\sqrt{\gamma^2 - 1}}{\gamma},
  \label{eq:vx_and_vtot}
\end{equation}
from which
\begin{equation}
  \cos\theta_i(z,t) = \sqrt{1 - \frac{1}{\gamma^2 - 1}\left(\frac{eB_y(z,t)d}{mc}\right)^2}.
  \label{eq:cos_i}
\end{equation}

Upon using Eq.~\eqref{eq:KE_balance}, it is easy to recognize that setting the radical expression in
Eq.~\eqref{eq:cos_i} to zero, i.e., $\gamma^2 - 1 - \left(eB_y d/mc\right)^2 = 0$, amounts to precisely
the Hull condition\footnote{The Hull condition can be derived from the same equations by
  setting $v_x=0$ at $x=d$, i.e., the condition that electrons do not reach the anode.}
\eqref{eq:Lovelace_Ott}. This means that the angle of incidence \eqref{eq:cos_i}
is undefined at insulation (by Lovelace and Ott) as well as for larger values of the magnetic field,
for which the expression under the square root becomes negative. 
In other words, using a Hull curve with a ``spillover'' is incompatible with the so outlined method of
calculating the temperature.
The simplest approach to resolving this incompatibility is to neglect temperature
contributions at locations where $B_y\ge B_{\rm crit}$. This simple solution is
acceptable for two reasons. First, we have seen from the Hull curve in Fig.~\ref{fig:hull_curve}
that the electron flux, which is proportional to the current,
significantly drops near $B_{\rm crit}$, and so does its contribution to the
temperature rise. And second, the neglected range of magnetic fields is only a small fraction of
$3$--$4$\% of the total range for our Analytic fit.
Our practical experience has confirmed that temperature doesn't noticeably increase at locations
where the magnetic field has exceeded the critical magnetic field.

\begin{figure}[t!!]
  \centering
  \includegraphics[width=0.4\textwidth]{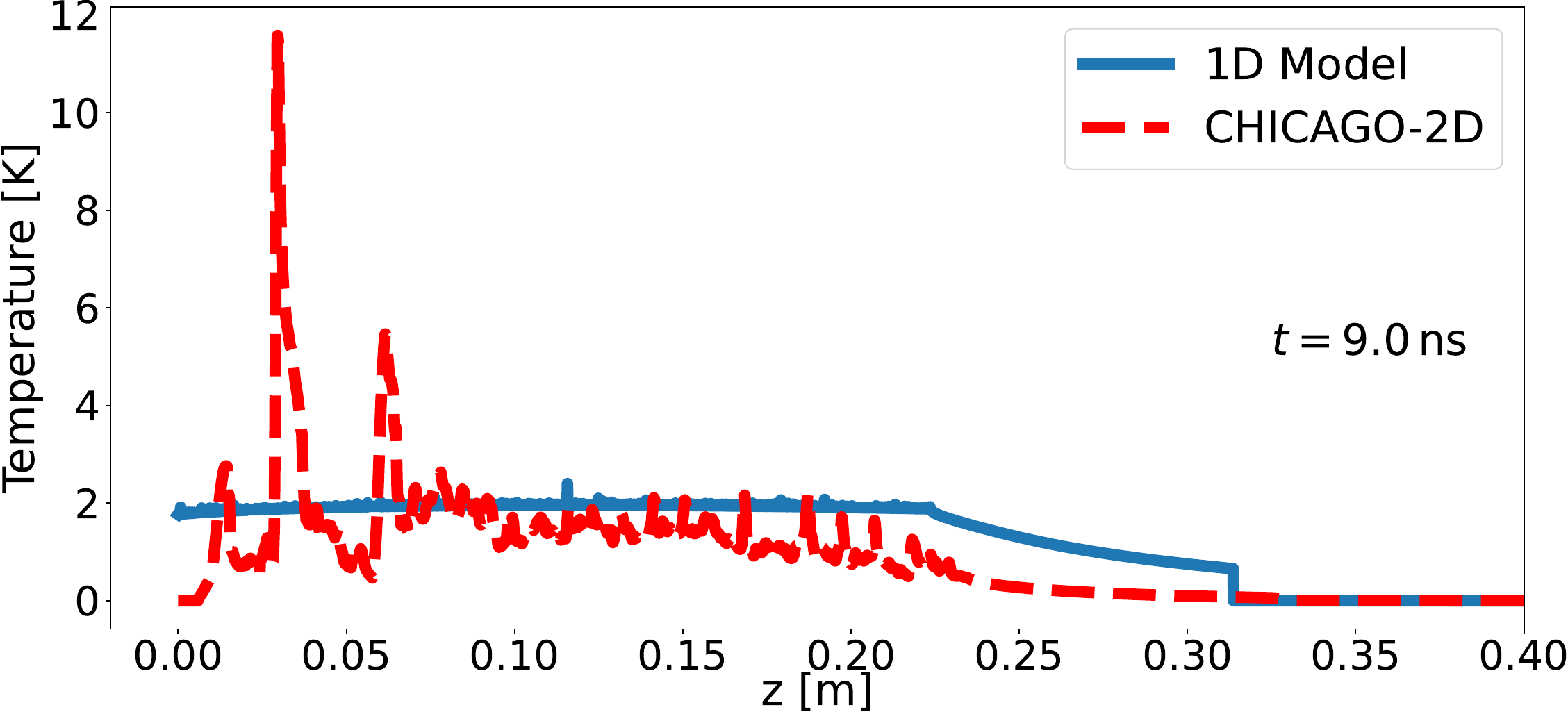} \\
  \includegraphics[width=0.4\textwidth]{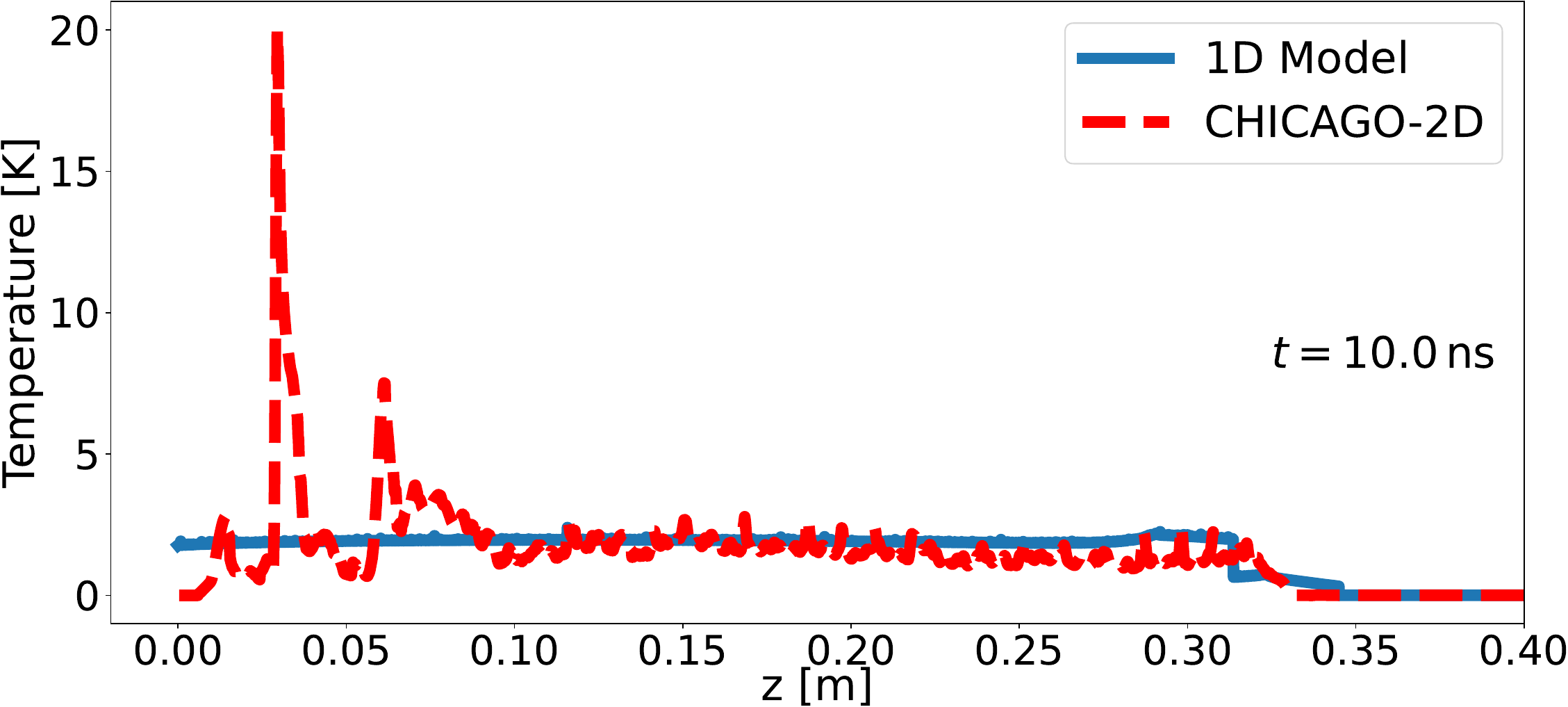}
  \caption{Comparison of temperature increase at $t=9\,$ns (top) and  $10\,$ns (bottom)
    between the 1D model and CHICAGO 2D PIC simulations,
    for $\Ipeak=20\,$MA/m, $\Taupeak=100\,$ns, AK gap, $d=5\,$mm, and $L=1\,$m.}
\label{fig:temperature}
\end{figure}

As a verification of this approach, Fig.~\ref{fig:temperature} 
shows the temperature rise calculated from data from our 1D model
and from CHICAGO 2D PIC, for aluminum metal initially at $300\,$K.
The simulation setup is the same as in Fig.~\ref{fig:SCL_current_densities}.
To avoid small differences due to algorithmic implementation details, we use CHICAGO's
output quantities of current density and kinetic energy near the anode
(quantities from particles information accumulated to the grid)
to perform the calculations leading to Eq.~\eqref{eq:temperature}. 
We notice that the overall agreement between the two models is excellent.
This implies also that not only do both components of the current density
agree very well (we have not shown $z$-components from either model) but so do 
the kinetic energies. The slight disagreement in the range of $20$--$30\,$cm at $t=9\,$ns (top panel)
is due to the slight temporal delay of the PIC pulse compared to the 1D model,
see Fig.~\ref{fig:SCL_current_densities} (bottom panel);
however, just $1\,$ns later, at $10\,$ns (bottom panel) no such disagreement
is seen (the current loss pulses have caught up to one another).
Since the temperature calculation is cumulative,
such temporal offsets would not affect the final temperature increase.

The obvious question concerns the large temperature oscillations
in the PIC simulation in the first $\sim 10\,$cm of the MITL.
These oscillations correspond to the current (and flux) oscillations in Fig.~\ref{fig:SCL_current_densities},
which we believe to be a numerical artifact of the SCL emission algorithm in PIC,
as discussed in Ref.~\cite{evstatiev_efficient_2023}.

\section{Conclusions}
\label{sec:conclusions}

In this work we have discussed MITL losses in a strongly time-dependent setting.
The physics that helps dramatically reduce current losses in long MITLs
is that of the diamagnetic, non-local, cumulative electromagnetic effect of SCL currents.
In short MITLs these effects diminish and in the limit vanish,
and current losses can be predicted by the Child-Langmuir SCL law in
the external (vacuum) EM fields.
By comparing quantities of merit in systems with strong time dependence
vs. those in (quasi-) steady-state, such as charge and current densities, current losses, etc.,
a more accurate definition of a long MITL is proposed.

A new 1D electromagnetic model is developed that performs on par with
circuit element codes,
with six to seven orders of magnitude lower computational cost than 2D PIC computations.
The model is verified against 2D PIC simulations
in a parallel plate Cartesian geometry as well as in a straight coaxial MITL geometry
with \tcb{azimuthal} symmetry. This allows for efficient navigation of the large MITL design parameter space.
A number of scaling studies have been performed, helping to understand and
minimize MITL losses. A temperature diagnostic within the 1D model is easy to implement,
based on the available information from the fields and currents.
Agreement with the temperature calculations from PIC data further verifies
that the 1D model accurately predicts kinetic energies and electron incidence angles as well.

A revised magnetic insulation model is proposed in the form a Hull curve,
partly informed by direct PIC simulations and partly by fitting PIC simulation data.
Major circuit element codes using a previously computed Hull curve
tend to overpredict current losses in long MITLs but implementation of our
revised Hull curve should not present difficulty. 

It is proposed that previous experimental observations
by Orzechowski and Bekefi of non-zero loss current beyond the theoretical
critical magnetic field by Lovelace and Ott is due to the electron
layer losing stability, not due to the generation of hot electrons,
a physics picture consistent with our PIC simulations.
This hypothesis still needs experimental confirmation.

\tcg{Several extensions to our 1D model could be explored in the future.
  Including the transient motion of electrons across the AK gap
  could be achieved following calculations similar to those in
  Ref.~\cite{wong_multipactor_2019}.}
\tcg{Use of the relativistic Child-Langmuir law by Jory and Trivelpiece
  \cite{jory_relativistic_1969} should be straightforward, although the connection
  \eqref{eq:E_tot} would need revision.}
\tcr{The inclusion of bipolar flow is another important extension left for the future.
  Thermally desorbed ions are frequently encountered in radially converging geometries
  and in many situations their contribution to current losses cannot be neglected.}
Thus, another extension to our model is to non-trivial geometries.
Such opportunity is presented within our model by using geometrical
transformations of fields and sources, for which well established techniques exist.
Important geometric effects not captured by our 1D model\,---\,as well as by CEM models\,---\,are
field enhancement around corners and the more general transverse magnetic (TM) modes in MITLs.

\vspace{-10pt}

\acknowledgments

The authors acknowledge useful feedback and discussions
with Greg Frye, Josh Leckbee, Keith Matzen, Kate Bell,
David Sirajuddin and Christopher Jennings from SNL.
Sandia National Laboratories is a multi-mission laboratory managed and
operated by National Technology \& Engineering Solutions of Sandia,
LLC (NTESS), a wholly owned subsidiary of Honeywell International
Inc., for the U.S. Department of Energy’s National Nuclear Security
Administration (DOE/NNSA) under contract DE-NA0003525. This written
work is authored by an employee of NTESS. The employee, not NTESS,
owns the right, title and interest in and to the written work and is
responsible for its contents. Any subjective views or opinions that
might be expressed in the written work do not necessarily represent
the views of the U.S. Government. The publisher acknowledges that the
U.S. Government retains a non-exclusive, paid-up, irrevocable,
world-wide license to publish or reproduce the published form of this
written work or allow others to do so, for U.S. Government
purposes. The DOE will provide public access to results of federally
sponsored research in accordance with the DOE Public Access Plan.
This work was funded by Laboratory Directed Research and Development
grant No.~229292.


\section{Data availability}
Data may be available upon reasonable request.

\section{Author declarations}
The authors have no conflicts to disclose.

\numberwithin{equation}{section}

\begin{appendices}

  \section{Coaxial geometry}
  \label{sec:coaxial_MITL}

  A straight coaxial MITL can be simulated within the 1D model with minor adjustments.
  Calculations in the cylindrical geometry $(r,\theta,z)$ with azimuthal symmetry
  (in the $\theta$-direction) closely parallel those in Cartesian.
  For our purposes, invoking the small AK gap approximation \ref{assumption:AK_gap}
  allows to consider the radius of the coaxial MITL as only a parameter.
  Within the error of that approximation, that radius can be chosen as either the inner or the outer.
  Our choice of cathode is typically the inner electrode, while the anode is the outer one.
  Accordingly, we choose the outer radius when comparing fields and current densities at the anode.

  We can make the following adjustments in the 1D model. Denoting the radius of the coaxial MITL by $R$,
  we adjust the current density in the wave equation \eqref{eq:wave_eq} to
  \begin{equation}
    j_x \longrightarrow j_r  = \frac{j_x}{2\pi R}
    \label{appeq:j_r}
  \end{equation}
  The electric and magnetic fields follow from the same Eqs.~\eqref{eq:E_x}, \eqref{eq:B_y}
  but with the correction
  \begin{equation}
    E_x \longrightarrow \frac{E_x}{2\pi R}, \qquad B_{\theta} = \frac{E_r}{c}.
    \label{appeq:E_B}
  \end{equation}

  Intuitively, in the small AK gap approximation the coaxial MITL is expected to behave as Cartesian.
  Our simulations have confirmed that such an adjusted ``coaxial'' 1D model also has excellent
  agreement with both CHICAGO 2D PIC simulations (in the $(r,z)$ coordinates)
  and EMPIRE PIC simulations of a thin 3D wedge.

\end{appendices}


\bibliographystyle{unsrt}

\end{document}